\documentclass[aps,prl,twocolumn,superscriptaddress,floatfix]{revtex4-1}
\usepackage{color}
\usepackage{array}
\usepackage{verbatim}
\usepackage{multirow}
\usepackage{amsmath}
\usepackage{amssymb}
\usepackage{dsfont}
\usepackage{graphicx}
\usepackage{threeparttable}
\usepackage{esint}
\usepackage[unicode=true,
 bookmarks=true,bookmarksnumbered=true,bookmarksopen=false,
 breaklinks=true,pdfborder={0 0 0},backref=false,colorlinks=true,
urlcolor=blue,linkcolor=blue,citecolor=blue] {hyperref}%
\usepackage{cleveref}
\usepackage{float}
\usepackage{longtable}
\usepackage{diagbox}
\usepackage{tabularx}
\usepackage{setspace}
\begin{document}
\title{Group theory of Raman effect in magnetic materials}
\author{Rui-Chun Xiao}  \email{xiaoruichun@ahu.edu.cn}
\affiliation{Institute of Physical Science and Information Technology, Anhui University, Hefei 230601, China}
\affiliation{Anhui Provincial Key Laboratory of Magnetic Functional Materials and Devices, School of Materials Science and Engineering, Anhui University, Hefei 230601, China}
\author{Xue Liu}  %lxue@ahu.edu.cn
\affiliation{Center of Free Electron Laser \text{\&} High Magnetic Field, Anhui University, Hefei 230601, China}

\author{Yuxuan Jiang} %yuxuan.jiang@ahu.edu.cn
\affiliation{School of Physics, Anhui University, Hefei 230601, China}

\author{Hang Zhou}
\affiliation{Key Laboratory of Materials Physics, Institute of Solid State Physics, Hefei Institutes of Physical Science, Chinese Academy of Sciences, Hefei 230031, China}

\author{Zi-Hao Feng}
\affiliation{Institute of Physical Science and Information Technology, Anhui University, Hefei 230601, China}

\author{Jie Hou}
\affiliation{Institute of Physical Science and Information Technology, Anhui University, Hefei 230601, China}

\author{Xiangru Kong} % kongxiangru@mail.neu.edu.cn
\affiliation{College of Sciences, Northeastern University, Shenyang, 110819, China}

\author{Yujun Zhang} % zhangyujun@ynu.edu.cn
\affiliation{School of Physics and Astronomy and Key Lab of Quantum Information of Yunnan Province, Yunnan University, Kunming 650091, China}

\begin{abstract}
 Despite the wealth of experimental observations on Raman scattering in magnetic materials, the underlying selection rules have remained largely unexplored. In this work, we use Onsager reciprocity relation, other than the conventional corepresentation method, to deal with the mathematical structures of Raman tensors in magnetic groups. Using this approach, we generate Raman tensor tables for all magnetic point groups, and present a comprehensive understanding of the Raman selection rules in magnetic materials with direct product representations method. Our theoretical and numerical results match previous experiments well, and resolve a recent puzzle in the Raman spectroscopy of CrSBr. Moreover, we identify a common but overlooked phenomenon: the magneto-Raman vector can be orthogonal to the magnetic moment direction. Our method and associated Raman tensor tables will be helpful for the Raman studies in both experimental and theoretical domains.
\end{abstract}

\maketitle

%===========================================================%
\textbf{1. Introduction}

\begin{figure}[!htbp]
\centering
\includegraphics[width=0.45\textwidth]{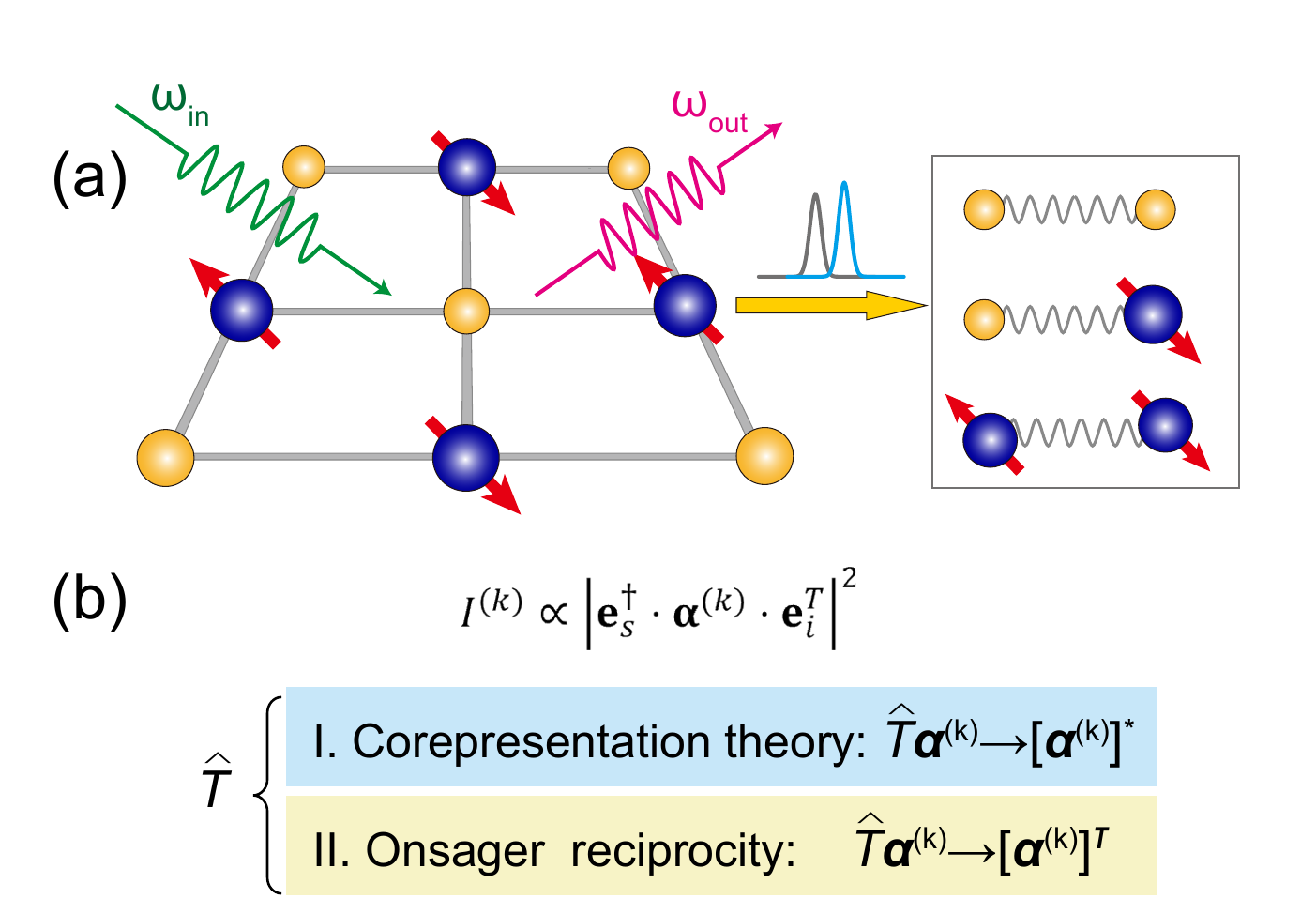}
\caption{(a) Schematic illustration of Raman scattering in magnetic materials, enabling the detection of lattice vibrations and spin-phonon excitations. (b) Two methods for imposing time-reversal symmetry constraints on the Raman tensor: (I) adopted in previous studies \cite{RN4022, RN4026}, and (II) adopted in this work.%: (1) complex conjugation in the conventional corepresentation theory, and (2) transposition via the Onsager reciprocal relation adopted in this work.
}
\label{Fig1}
\end{figure}

 Raman scattering is a powerful probe of excitations in condensed matter through inelastic light scattering. In magnetic materials, Raman spectroscopy provides unique insights into lattice vibrations (phonons), spin excitations (including magnons), and their mutual interactions [Fig. \ref{Fig1}(a)]. Similar to that of nonlinear optics \cite{RN2808, RN3052, RN3668, RN3814, RN3841, RN4135}, and magneto-optics effect \cite{RN2479, RN3871, RN3793, RN4045, RN3940,RN4079}, many intriguing magneto-Raman effects have been discovered in magnetically ordered materials recently. For example, the magneto-Raman \cite{RN4012, RN4011, RN4025, RN4010, RN3400,RN4031} can arise in both ferromagnetic (FM) and antiferromagnetic (AFM) materials. The odd-parity ($u$) phonons, which are forbidden in the paramagnetic state, can be Raman active in $PT$-symmetric AFM (such as CrI$_3$ \cite{RN4012, RN4011, RN4025}). The doubly degenerate $E$ modes can be split into left- and right-handed chiral phonons in FM CrBr$_3$ \cite{RN4017} and Co$_3$Sn$_2$S$_2$ \cite{RN3942, RN4032}. Even though progress has been made in experiments, a unified framework on the Raman selection rules is yet to be fully established.

% The selection rules of Raman scattering depend on the Raman tensors, which are constrained by the symmetries of the materials, and serve as a guide for experimental detection.
The Raman tensors, dictated by the underlying crystal symmetries, govern the selection rules and thereby provide a roadmap for interpreting experimental spectra.
The Raman scattering tensor for nonmagnetic materials were systematically summarized with group theory \cite{RN4093, RN4041}, and have been successfully applied in experiments. In this method, the Raman tensors are viewed as a three-dimensional representation of a specific point group. However, the Raman selection rules and associated tensors in magnetically ordered materials are more complex than in nonmagnetic cases \cite{RN4012, RN4011, RN4025, RN4010}, mainly due to the constraints of time-reversal ($T$) and other anti-unitary symmetry ($R'$).
Since a specific Raman tensor form irreducible representation (irreps) of a nonmagnetic group, one might naturally expect them to serve as corepresentations of the magnetic groups \cite{RN1704}, wherein the symmetry constraints imposed by time-reversal on them manifest complex conjugate operation [method I in Fig. \ref{Fig1}(b)]. Following this conception, Cracknell \cite{RN4022} and Anastassakis \cite{RN4026} \emph{et.al.} made early attempts to derive Raman tensors for magnetic groups from the 1960s to the 1970s. However, their theory has not gained experimental validation in recent studies \cite{RN4012, RN4011, RN4025, RN4010}.

In the 1980s-1990s, studies on the non-equilibrium response tensors found that the dissipative process under time reversal follow Onsager reciprocal relation \cite{RN4105, RN4089, RN4087, RN4086, RN4088}, rather than complex conjugation. Given that the Raman effect is an inelastic scattering process, it also exhibits non-equilibrium properties. Consequently, the time reversal constraint on Raman tensors should follow the Onsager reciprocal relation [method II in Fig. \ref{Fig1}(b)], which offers a crucial direction for this long-standing issue.

%
% the Raman tensor is the derivative of the dielectric response tensor with respect to phonon coordinates. Given that the dielectric response obeys the Onsager reciprocal relation, the Raman tensor naturally inherits this constraint under time reversal, providing a crucial direction for further investigation.

In this work, we employ Onsager reciprocal relation to address the time reversal and anti-unitary operations constrained on Raman tensors. Using this approach, we establish comprehensive Raman tensor tables for all magnetic point groups (MPGs). We further investigate the mathematical structure of the Raman tensor in MPGs by means of direct product representations, thereby clarifying Raman selection rules in magnetic materials. Our symmetry analysis is well consistent with previous experimental results (\emph{e.g.}, CrI$_3$ \cite{RN4012, RN4011, RN4025} and CrSBr \cite{RN4146, RN4038, RN4042}).

\textbf{}

\textbf{2. Method and results}

\textbf{2.1 The symmetry-constrained Raman tensor in all MPGs}

{In nonmagnetic point groups, the Raman tensors transform according to the irreducible representations (irreps) of the corresponding point group.}
To be specific, for a symmetry operation $R$ belonging to group $G$, the Raman tensors ${{\boldsymbol{\alpha }}^{(k)}}$ ($k$=1,2,...,$d$) of the $r$-th phonon mode with degeneracy $d$ must satisfy the following transformation rule \cite{RN4093, RN4041}:
\begin{equation}
	\mathbf{D}(R){{\boldsymbol{\alpha }}^{(k)}}\mathbf{D}{{(R)}^{T}}=\sum\limits_{p=1}^{d}{\Gamma _{pk}^{(r)}(R){{\boldsymbol{\alpha }}^{(p)}}},
\label{Eq-Raman_1}
\end{equation}
where $\mathbf{D}\left( R \right)$ is the operation matrix of $R$, and $\boldsymbol{\Gamma}^{(\mathrm{r})}({R})$ is the $r$-th irrep matrix. The left side represents a transformation applied to a single Raman tensor, while the right side denotes a transformation between tensors.

Eq. (\ref{Eq-Raman_1}) is complicated due to the involvement of multiple indices. However, for the one-dimensional (1D) phonon mode ($d$=1), the representation matrix $\Gamma^{({r})}({R})$ reduces to the character $\chi^{({r})}({R})$, simplifying the expression to:
\begin{equation}
  \mathbf{D}(R)\boldsymbol{\alpha D}{{(R)}^{T}}={{\chi }^{(r)}}(R)\boldsymbol{\alpha }.
\label{Eq-Raman_2}
\end{equation}
Eq. (\ref{Eq-Raman_2}) resembles the symmetry constrained 2\textsuperscript{nd} order response tensor (such as dielectric function tensor or conductivity), except for the character $\chi^{(\mathrm{r})}({R})$ on the right.

Besides, the time-reversal symmetry enforces that the Raman tensor has the interchange symmetry of the indices ($ij$ $\leftrightarrow$ $ji$). With the above symmetry constraint, the Raman tensors for all non-magnetic point groups can be found in textbooks, the literature \cite{RN4041}, and online databases \cite{RN4090}.

When the system lacks time-reversal symmetry, the unitary ($R$) and antiunitary ($R'=TR$) operations within an MPG $M$ impose distinct constraints on the Raman tensor. As noted in the introduction, $T$ constraints on ${{\boldsymbol{\alpha }}^{(k)}}$ should obey the Onsager reciprocal relation. Therefore, following the treatment of electrical and optical responses in magnetic systems \cite{RN4089, RN4087, RN4086, RN4088, RN3940}, for an antiunitary operation $R'$, the spatial part $R$ imposes constraints analogous to those in Eq. (\ref{Eq-Raman_1}), while the time-reversal operator $T$ exchanges tensor indices ($ij$ $\leftrightarrow$ $ji$). Thus, the overall constraint on the Raman tensor is:
\begin{equation}
\left\{
\begin{aligned}
& R:\sum_{mn} R_{im} R_{jn} \alpha_{mn}^{(k)} = \sum_{p=1}^d \Gamma_{pk}^{(r)}(R) \alpha_{ij}^{(p)}, \\
& R':\sum_{mn} R_{im} R_{jn} \alpha_{mn}^{(k)} = \sum_{p=1}^d \Gamma_{pk}^{(r)}(R) \alpha_{ji}^{(p)}.
\end{aligned}
\right.
\label{Eq-Raman_3}
\end{equation}
where $R_{ij}$ is the element of $~\mathbf{D}\left( R \right)$ in Eq. (\ref{Eq-Raman_1}).

To better investigate the influence of MPG $M$ on the Raman tensor $\left[\alpha_{i j}\right]$, we decompose it into symmetric part $\left[\alpha_{i j}^S\right]$ and antisymmetric part $\left[\alpha_{i j}^A\right]$:
\begin{equation}
	\left[ {{\alpha }_{ij}} \right]=\left[ \begin{array}{*{35}{l}}
   \alpha _{xx}^{S} & \alpha _{xy}^{S} & \alpha _{xz}^{S}  \\
   \alpha _{xy}^{S} & \alpha _{yy}^{S} & \alpha _{yz}^{S}  \\
   \alpha _{xz}^{S} & \alpha _{yz}^{S} & \alpha _{zz}^{S}  \\
\end{array} \right]+\left[ \begin{matrix}
   0 & \alpha _{xy}^{A} & -\alpha _{zx}^{A}  \\
   -\alpha _{xy}^{A} & 0 & \alpha _{yz}^{A}  \\
   \alpha _{zx}^{A} & -\alpha _{yz}^{A} & 0  \\
\end{matrix} \right].
\label{Eq-Raman_4}
\end{equation}

Based on Eqs. (\ref{Eq-Raman_3}) and (\ref{Eq-Raman_4}), the constraint on the Raman tensors $\left[\alpha_{i j}^S\right]$ and $\left[\alpha_{i j}^A\right]$ can be written as:
\begin{equation}
\left\{
\begin{aligned}
  & R:\sum\limits_{mn}{{{R}_{im}}}{{R}_{jn}}\alpha _{mn}^{(k)}=\sum\limits_{p=1}^{d}{\Gamma _{pk}^{(r)}(R)\alpha _{ij}^{(p)}}, \\
 & R':\sum\limits_{mn}{{{R}_{im}}}{{R}_{jn}}\alpha _{mn}^{(k)}=\pm \sum\limits_{p=1}^{d}{\Gamma _{pk}^{(r)}(R)\alpha _{ij}^{(p)}.} \\
\end{aligned}
\right.
\label{Eq-Raman_5}
\end{equation}
The +/$-$ symbol in the second equation corresponds to the transformation of the symmetric/anti-symmetric part under the anti-unitary operation $R'$.

Based on the above method, we wrote a computational code to determine the forms of the Raman tensors for all irreps under each MPG. In the Supplementary materials (Section I), we summarize the Raman tensors in tabulated form, which is also made available on a dedicated website [\href{https://ruichun.github.io/MagneticRamanTensor/} {https://ruichun.github.io/MagneticRamanTensor/}] for convenient use by researchers. {Brief introductions to the corepresentation method and phenomenological model method are given in Section II of the Supplementary materials, where their differences with our method are also listed.}

Taking bilayer CrI$_3$ as an example, we find that the Raman tensors (Table \ref{Tab1}) obtained by the above method for both FM ($2'/m'$) and AFM ($2/m'$) configurations are consistent with experimental results \cite{RN4012, RN4011, RN4025} and theoretical calculations  \cite{RN3941}. In contrast, the corepresentation method \cite{RN4022, RN4026} fails to describe the Raman behaviors in CrI$_3$.
Furthermore, we identify an additional antisymmetric tensor component $\alpha_{23}^A$ ($=-\alpha_{32}^A$)  for the $B_u$ mode in AFM state and $A_g$ phonon in the FM state, which has a small magnitude and cannot be captured by the previous phenomenological theory \cite{RN4011, RN3941}. This finding is further supported by our first-principles calculations (see Section VI of Supplementary materials).

% Furthermore, we identify an additional antisymmetric tensor component $\alpha_{23}^A$ ($=-\alpha_{32}^A$) for $B_u$ mode at AFM state and $A_g$ phonon at the FM state with a small magnitude, which cannot be captured by the phenomenological theory \cite{RN4011, RN3941}. This finding is further supported by our first-principles calculations, see Section V in Supplementary materials \cite{RN3775}.

\begin{table}[!htpb]
\centering
\caption{An example of the Raman tensors for $A_g$ and $B_u$ modes for bilayer CrI$_3$ based on our method. The first row lists [$\alpha_{ij}^S$], while the remaining rows correspond to [$\alpha_{ij}^A$]. The mapped irrep of each MPG is also indicated, whose meanings are given in Table \ref{Tab2}. }
\label{Tab1}
\resizebox{\columnwidth}{!}{%
\begin{tabular}{l|c|c}
\hline  \hline
$C_{2 h}(2 //y)$ & $A_g$ & $B_u$ \\
\hline
Symmetric part & $\left(\begin{array}{ccc}\alpha_{11} & 0 & \alpha_{13} \\ 0 & \alpha_{22} & 0 \\ \alpha_{31} & 0 & \alpha_{33}\end{array}\right)$ & 0 \\
\hline
\begin{tabular}{c}
Anti. Sym. of \\
$2 / {m}^{\prime}\left(AFM, {A}_{{u}}\right)$   \end{tabular} & 0 & $\left(\begin{array}{ccc}0 & A_{12} & 0 \\ -A_{12} & 0 & A_{23} \\ 0 & -A_{23} & 0\end{array}\right)$ \\
\hline
\begin{tabular}{c}
Anti. Sym. of \\
$2^{\prime} / {m}^{\prime}\left(FM, {B}_{{g}}\right)$  \end{tabular} & $\left(\begin{array}{ccc}0 & A_{12} & 0 \\ -A_{12} & 0 & A_{23} \\ 0 & -A_{23} & 0\end{array}\right)$ & 0 \\
\hline \hline
\end{tabular}
}
\end{table}
%%%%%%%%%%%%%%%%%%%%%%%%%%%%%%%%%%%%%%%%%%%%%%%%%%%%%%%%%%%%%%%%%%%%%%%%%%%%%%%%%

\textbf{2.2 The group structures of Raman tensors}

Although the above approach yields numerical results, an understanding of their group structure still requires elucidation. In the following, we reflect the above results in a profound physical meaning.
Based on Eqs. (\ref{Eq-Raman_3}) and (\ref{Eq-Raman_4}), the unitary and anti-unitary operations have the same constraint on $[\alpha_{ij}^S]$. Therefore, $[\alpha_{ij}^S]$ under the MPG $M$ is the same as it is in the isomorphic nonmagnetic point group $G$. Moreover, the mathematical structure $[\alpha_{ij}^S]$ is well understood in nonmagnetic point group $G$, \emph{i.e.}, the tensor elements contain the corresponding $r^2$ basis (such as $x^2$, $xy$, ... in Table \ref{Tab2}). Therefore, the key lies in analyzing the  antisymmetric part. Then, we analyze $[\alpha_{ij}^A]$ in the following five steps.

(I) According to Eq. (\ref{Eq-Raman_5}), the constraint of $R/R'$ on $[\alpha_{ij}^A]$

\begin{equation}
\sum\limits_{mn}{{{R}_{im}}}{{R}_{jn}}\alpha _{mn}^{A(k)}=\pm \sum\limits_{p=1}^{d}{\Gamma _{pk}^{(r)}(R)\alpha _{ij}^{A(p)}}.
\label{Eq-Raman_6}
\end{equation}
where the + and $-$ signs correspond to unitary $R$ and antiunitary $R'$ operations, respectively. % Apart from an additional minus sign, the action of $R'$ on $[\alpha_{ij}^A]$ resembles that of the unitary operation $R$.

(II) According to Eq. (\ref{Eq-Raman_4}), $[\alpha_{ij}^A]$ tensor contains only three independent components. By introducing the following substitution:
\begin{equation}
\alpha _{ij}^{A}={{\epsilon }_{ijk}}{{J}_{k}}, 	
\label{Eq-Raman_7}
\end{equation}
where ${{\epsilon }_{ijk}}$ is the Levi-Civita symbol. Therefore, $[\alpha_{ij}^A]$ can be mapped onto an axial vector $\mathbf{J}=({{J}_{x}},{{J}_{y}},{{J}_{z}})$. Consequently, based on Eqs. (\ref{Eq-Raman_6}) and (\ref{Eq-Raman_7}), the transformation of $\left[\alpha_{i j}^{A(k)} \right]$ tensors under $R /R'$  reduces to:
\begin{equation}
	\sum\limits_{m}{{{R}_{im}}}J_{m}^{(k)}=\pm \sum\limits_{p=1}^{d}{\Gamma _{pk}^{(r)}(R)J_{i}^{(p)}}.
\label{Eq-Raman_8}
\end{equation}

(III) Since $R$ and $R'$ are elements of the black-and-white MPG $M$ \cite{RN4098}, we may assign them the values +1 and $-1$, respectively. In this way, the MPG $M$ can be regarded as a 1D real irrep ${{\chi }_{mag}}$ of the isomorphic nonmagnetic point group $G$. As shown in Table \ref{Tab2}, for the MPG $2'/m'$, the unitary operations $E$ and $\bar{1}$ are assigned with +1, while $2'$ and $m'$ correspond to $-1$, leading to this MPG with $B_g$ irrep of $2/m$. Similarly, the MPG $2/m'$ corresponds to $A_u$ irrep of $2/m$.

(IV) Consequently, Eq. (\ref{Eq-Raman_8}) can be further written as:
\begin{equation}
	\sum\limits_{m}{{{R}_{im}}}J_{m}^{(k)}={{\chi }_{mag}}(R)\sum\limits_{p}{\Gamma _{pk}^{(r)}}(R)J_{i}^{(p)}.
\label{Eq-Raman_9}
\end{equation}
According to the representation theory, the multiplication of the irrep ${{\chi }_{mag}}$ and phonon mode $\Gamma^{(r)}$ yields a direct product representation ${{\chi }_{mag}}\otimes \Gamma _{{}}^{(r)}$, which is also an irrep of $G$.

(V) Assuming ${{\chi }_{mag}}\otimes \Gamma _{{}}^{(r)}={{\Gamma }^{(a)}}$, Eq. (\ref{Eq-Raman_9}) can be rewritten as:
\begin{equation}	
\sum\limits_{m}{{{R}_{im}}}J_{m}^{(k)}=\sum\limits_{p}{\Gamma _{pk}^{(a)}}(R)J_{i}^{(p)}.
\label{Eq-Raman_10}
\end{equation}	
Thus, Eq. (\ref{Eq-Raman_10}) implies that if the irrep $\Gamma^{(a)}$ permits axial vectors $\mathbf{J}^{(k)}$, then the phonon mode $\Gamma _{{}}^{(r)}$ under the MPG $M$ (${{\chi }_{mag}}$) exhibits magneto-Raman tensor $[\alpha_{ij}^A]$.

\begin{figure}[!htbp]
\centering
\includegraphics[width=0.49\textwidth]{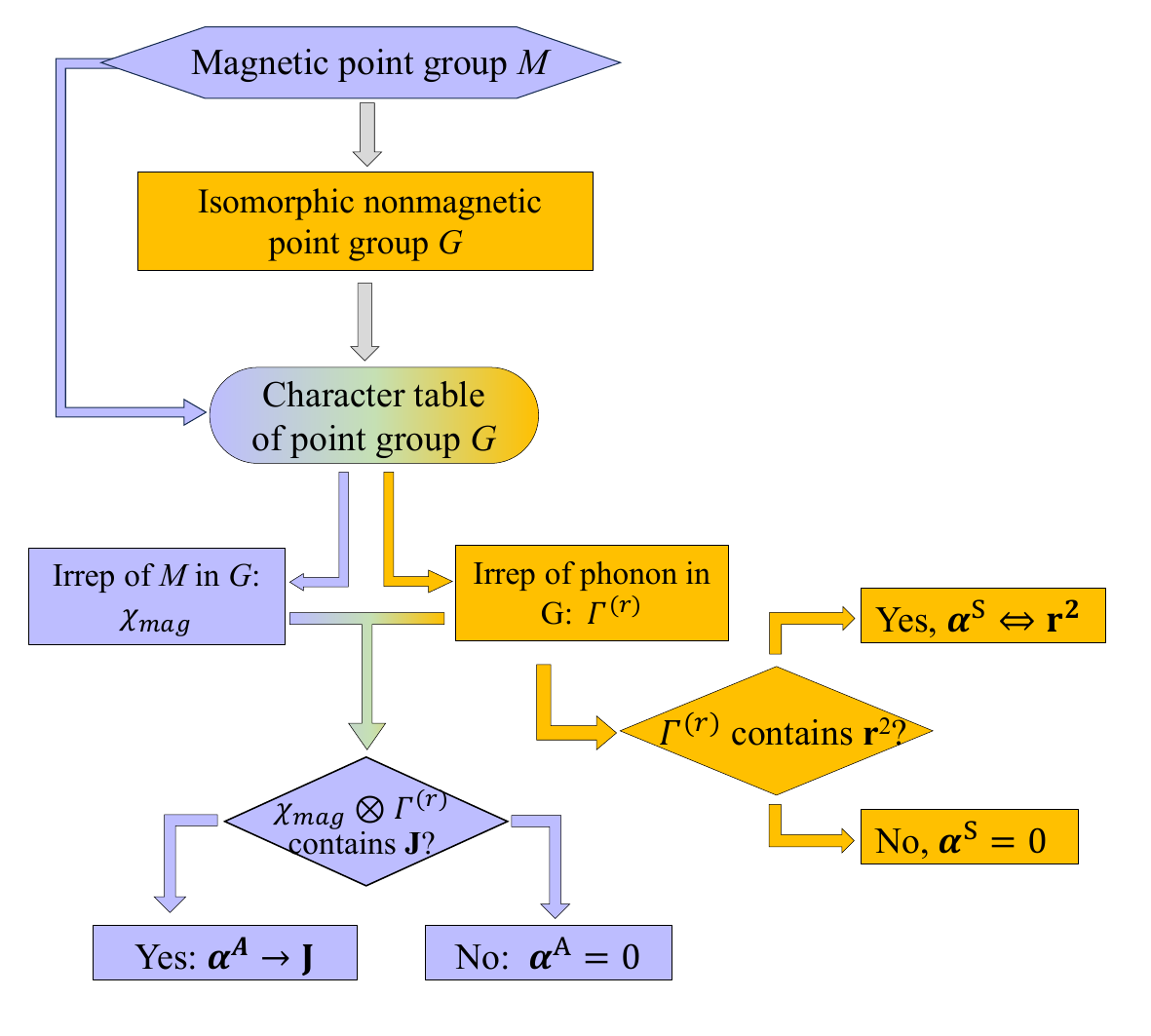}
\caption{%Schematic of the group representation theory method used to determine the Raman tensor for a given MPG. The symmetric part of the Raman tenors can be determined by the basis of $\Gamma^{(r)} $. For the antisymmetric Raman tensor, it is determined by whether the direct product representation $\chi_{\text {mag }} \otimes \Gamma^{(r)}$ permits the existence of axial vectors $\mathbf{J}$. This method requires no numerical computation and clearly reveals the group structure.
Schematic of the group-theoretical method for determining the Raman tensor for a given MPG. The symmetric part is obtained from the basis of $\Gamma^{(r)}$, while the antisymmetric part is determined by whether the direct product representation $\chi_{\text {mag }} \otimes \Gamma^{(r)}$, allows axial vectors $\mathbf{J}$. This method requires no numerical computation and clearly reveals the group structure.
}
\label{Fig2}
\end{figure}

Based on the above analysis, the group structure of the Raman tensor can be clearly understood. The logical framework of the method is illustrated in Fig. \ref{Fig2}. Consequently, analogous to the determination of symmetric Raman tensors in nonmagnetic materials, all Raman tensor components in $M$ can be obtained solely from the point group character table of $G$ that includes basic basis functions.

\begin{table}[!htpb]
\centering
\caption{Character table of $C_{2h}$  (2//$y$), together with the corresponding MPGs and basis functions. Using this table, both $[\alpha_{ij}^S]$ and $[\alpha_{ij}^A]$ can be determined.}
\label{Tab2}
\resizebox{0.95\columnwidth}{!}{
\begin{tabular}{c|l|l|l|l|l|l|l|l}
\hline \hline
${C}_{2h}(2 / {m})$ & $E$ & 2 & $\overline{1}$ & $m$ & \multicolumn{3}{c|}{Basis} & MPG \\
\hline ${A}_{{g}}$ & 1 & 1 & 1 & 1 &  & ${x}^2, {y}^2, {z}^2, {xz}$ & $J_y$ & 2/m \\
\hline ${B}_{{g}} $ & 1 & -1 & 1 & -1 & & $xy$, $yz$ & $J_x$, $J_z$ & $2^{\prime} / {m}^{\prime}$ \\
\hline ${A}_{{u}}$ & 1 & 1 & -1 & -1 & $y$ & & & $2 / {m}^{\prime}$ \\
\hline ${B}_{{u}}$ & 1 & -1 & -1 & 1 & $x$, $z$ & & & $2'/{m}$ \\
\hline \hline
\end{tabular}
}
\end{table}

We apply our method to bilayer CrI$_3$ again. The quadratic basis functions of the ${A}_{{g}}$ mode are ${x}^2, {y}^2, {z}^2, {xz}$ according to Table \ref{Tab2}; therefore the $\left[\alpha_{{ij}}^{{S}}\right]$ of $A_g$ phonon has the corresponding counterpart. The MPG of bilayer CrI$_3$ in FM state is $2^{\prime} / {m}^{\prime}$, which corresponds to ${B}_{{g}}$ irrep of the ${C}_{2h}$ point group. Because $A_g \otimes B_g=B_g$, ${B}_{{g}}$ permits axial vectors ${J}_{{x}}$ and ${J}_{{z}}$, according to Table \ref{Tab2}. Consequently the Raman tensor of this ${A}_{{g}}$ phonon mode under the MPG $2^{\prime} / {m}^{\prime}$ takes the same form as shown in Table \ref{Tab1}. Similarly, the $B_u$ phonon in the AFM state ($A_u$) exhibits magneto-Raman activity, with $[\alpha_{ij}^A]$ mapping into ${J}_{{x}}$ and ${J}_{{z}}$.
%Similarly, in the AFM state of bilayer CrI$_3$, the MPG is $2/m'$, which corresponds to the $A_u$ irrep of $C_{2h}$. For the $B_u$ phonon mode, ${{A}_{u}}\otimes {{B}_{u}}={{B}_{g}}$, and $B_g$ permits axial vectors. Therefore, the $B_u$ mode, which is Raman inactive in the nonmagnetic state (no  ${{r}^{2}}$ terms), becomes active in the presence of magnetism. These results are consistent with the purely mathematical results presented in Table \ref{Tab2}.

According to the Placzek method \cite{RN4092} [$\alpha_{i j}^{(k)}(\omega)=\left.\frac{\partial \varepsilon_{i j}(\omega)}{\partial Q_k}\right|_{Q=0}$] and Eq. (\ref{Eq-Raman_4}), the magneto-Raman effect can be viewed as a phonon-assisted magneto-optical effect (refers specifically to the Faraday effect and Kerr effect here).
Given that $[\alpha_{ij}^A]$ originates from the direct product of $\Gamma^{(r)}$ and $M$ (${{\chi }_{mag}}$), a novel geometric relationship between the magnetic order (such as magnetic moment $\mathbf{m}$ for FM and Ne\'{e}l vector $\mathbf{n}$ for AFM materials) and the magneto-Raman tensor mapped axial vector $\mathbf{J}$ can emerge. For instance, in contrast to the conventional case $\mathbf{J} // \mathbf{m}$, the orthogonal configuration $\mathbf{J} \bot \mathbf{m}$ is also possible. Below, we use CrSBr as an example to demonstrate this relationship.

%%%%%%%%%%%%%%%%%%%%%%%%%%%%%%%%%%%%%%%%%%%%%%%%%%%%%%
\textbf{}

\textbf{2.3 A specific example on $\mathbf{J} \bot \mathbf{m} $}

\begin{figure}[!htbp]
\centering
\includegraphics[width=0.5\textwidth]{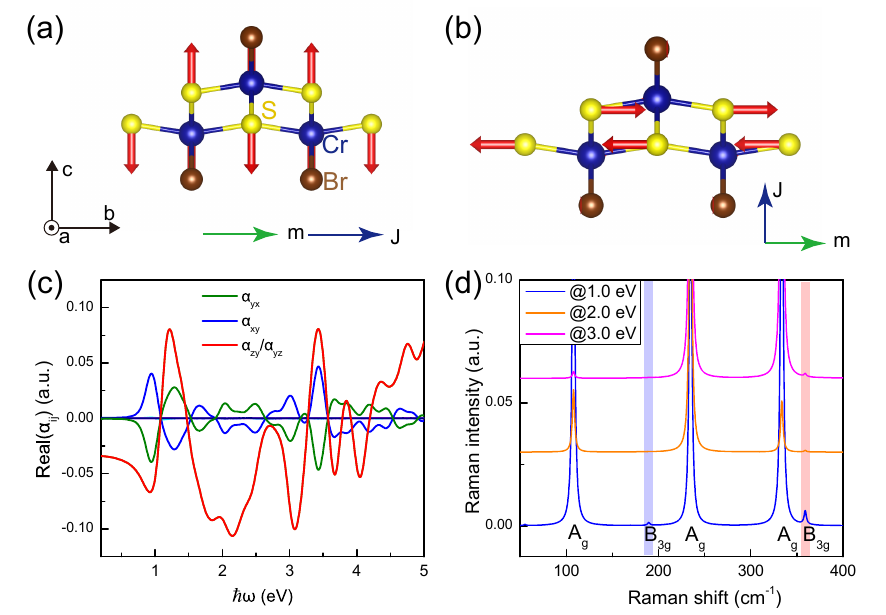}
\caption{ Calculated vibrational patterns of (a) $A_g$ (335 cm$^{-1}$) and (b) $B_{3g}$ (359 cm$^{-1}$) phonon modes for monolayer CrSBr. (c) Frequency-dependent real part Raman susceptibility of the $B_{3g}$ mode. (d) Simulated circularly polarized Raman spectra under the $\sigma^{-}/\sigma^{-}$ (left-left) configuration at different incident photon energies.}
\label{Fig3}
\end{figure}

CrSBr is a layered A-type AFM material, with the space group $Pmnm$. Below the Ne\'{e}l temperature, the magnetic atoms arrange in an interlayer AFM state, with magnetic moments aligning along the $b$-axis. In the paramagnetic state, the $A_g$, $B_{2g}$, and $B_{3g}$ phonon modes are Raman active.
In conventional experimental setups with incident light along $z$-direction, perpendicular to the sample, and only the $A_g$ modes [Fig. \ref{Fig3}(a)] are detectable \cite{RN3994, RN4091,RN4042}, because $B_{2g}$ and $B_{3g}$ have no in-plane Raman tensor elements in paramagnetic state. Nevertheless, a $B_{3g}$ mode [Fig. \ref{Fig3}(b)] at 355 cm$^{-1}$ was observed in few-layer CrSBr below the Ne\'{e}l temperature \cite{RN4146, RN4038, RN4042}, a feature that remained unexplained experimentally. This previously unresolved phenomenon is accounted for by our theory.

% We will explain this magneto-Raman mode with our symmetry theory and numerical calculation. We employ density functional perturbation theory (DFPT) or the frozen-phonon method to calculate the phonon frequencies at the $\Gamma$ point.
% The frequency-dependent dielectric function ${{\varepsilon }_{ij}}(\omega )$ is obtained based on the interband optical calculation \cite{RN3941}.
For the monolayer CrSBr with FM state, the MPG is $m' mm'$. According to Table S9 given in the Supplementary materials, the $B_{3g}$ phonon mode has the in-plane magneto-Raman elements due to the existing anti-symmetric Raman tensor elements ${{\alpha }_{12}}$ and ${{\alpha }_{21}}$ (${{\alpha }_{12}}=-{{\alpha }_{21}}$). This result is verified by our numerical calculation performed within the independent-particle resonant Placzek approach (see Section VII of Supplementary materials \cite{RN3941}), as shown in Fig. \ref{Fig3} (c).
Notably, at the zero frequency limit, only the symmetric parts of the Raman tensor are present, while the antisymmetric parts vanish. The behavior is consistent with the requirements of time-reversal symmetry.
Consequently, $B_{3g}$ modes are detectable in conventional experimental setups, as shown in Fig. \ref{Fig3}(d), which resolves the doubts raised in the recent experiment \cite{RN4146, RN4038, RN4042}.
More importantly, $[\alpha_{ij}^A]$ of $B_{3g}$ is mapped onto an axial vector $\mathbf{J}=(0,0,{{J}_{z}})$, which is perpendicular, rather than parallel, to the magnetic moment (along the $y$ axis), as illustrated in Fig. \ref{Fig3}(b). This unusual geometric relationship likely prevented previous studies from associating the
$B_{3g}$ mode with magnetism-induced Raman activity. This magneto-Raman mode can also be observed in the bilayer AFM state, but is absent in bulk (see Section VII of Supplementary materials). Another $B_{3g}$ mode at 189 cm$^{-1}$ possesses the same symmetry but exhibits considerably weaker Raman spectrum, as shown in Fig. \ref{Fig3}(d).

%%%%%%%%%%%%%%%%%%%%%%%%%%%%%%%%%%%%%%%%%%%%%%%%%%%
\textbf{}

\textbf{3. Discussion and conclusion}

Our study reveals that in many magnetic materials, even for antiferromagnets without magneto-optical effects, specific phonon modes can also exhibit a magneto-Raman effect. However, the observation may be hindered by weak spin-orbit coupling or insufficient spin-phonon coupling. Furthermore, we observe that the orthogonality between the magneto-Raman vector $\mathbf{J}$ and the magnetic order ($\mathbf{m}$ or $\mathbf{n}$) is quite general; see, for example, NiAs-type altermagnets in Supplementary materials.
{This geometric relation bears a strong resemblance to the relationship between the anomalous Hall vector (or magneto-optical vector) and the N\'eel vector in altermagnets \cite{RN4079,RN4045,RN3940}. A detailed analysis of this analogy is provided in Section IX of the Supplementary Material.}

{The materials studied in our work (such as CrSBr and bilayer CrI$_3$), are centrosymmetric. Consequently, geometric chiral phonons are forbidden by the preserved spatial inversion symmetry without magnetism, and circularly polarized Raman response can only originate from the broken time-reversal symmetry.} Moreover, the spectral characteristics of chiral phonons differ from those of magneto-Raman phonons, a point elaborated in Section IV of the Supplementary materials.

Based on the above theory, we can further develop frameworks such as Landau theory \cite{RN3736, RN3918} or the extrinsic parameter method \cite{RN3983, RN3940} to investigate the relationship between the Raman tensors and magnetic orders, which is valuable for determining the direction of magnetic moments using the Raman effect in experiments. Meanwhile, we can also employ spin groups \cite{RN3922, RN4102, RN4072} to explore the connection between Raman coefficients and the spin-orbit coupling (SOC)  effect. Given that magnons and phonons possess similar symmetry characteristics, our theoretical results are also applicable to Raman scattering involving magnons. Furthermore, this method is expected to apply to the infrared absorption spectra for magnetic materials.

In summary, we analyze the mathematical structures of the Raman tensors in magnetic systems with the group representation theory and Onsager reciprocal relations. By mapping the antisymmetric Raman tensor to an axial vector and leveraging the direct product of irreps, we conclude that the magneto-Raman effect can be viewed as a phonon-assisted magneto-optical effect. Our theory provides a good description of existing experiments (e.g., CrI$_3$) and offers a clear explanation for recent puzzling experimental observations (e.g., CrSBr). Furthermore, it establishes a unified framework for Raman selection rules, and the comprehensive Raman tensor tables can facilitate both theoretical and experimental research on Raman effect.

\textbf{}
\textbf{}

\textbf{Conflict of interest}
The authors declare that they have no conflict of interest.

\textbf{Acknowledgments}
This work is supported by the National Natural Science Foundation of China (Grants Nos. 12474100, 12204009, 12464002). We thank Tiantian Zhang, Bolin Li, Zeliang Sun, Gang Tang, Shu-Hui Zhang and Songming Wan for useful discussions on Raman experiments and theories.

\textbf{Author contributions}

Rui-Chun Xiao conceived and designed the project and carried out the symmetry analysis. Theoretical calculations were performed by Rui-Chun Xiao with assistance from Zi-Hao Feng, Jie Hou, and Hang Zhou. All authors contributed to the interpretation and discussion of the results. Rui-Chun Xiao drafted the manuscript and discussed with all co-authors.

\textbf{Appendix A. Supplementary materials}

Supplementary materials to this short communication can be found online at https://doi.org/****.

%%%%%%%%%%%%%%%%%%%%%%%%%%%%%%%%%%%%%
\bibliographystyle{apsrev4-1}
\bibliography{Reflatex}
%%%%%%%%%%%%%%%%%%%%%%%%%%%%%%%%%%%%
%%%%%%%%%%%%%%%%%%%%%%%%%%%%%%%%%%%%%%%%%%%%%%%%%%%%%%%%%

\clearpage

\begin{widetext}
\begin{center}
\begin{large}
\textbf{Supplemental Material for ``The group theory of Raman effect in magnetic materials"}
\end{large}
\end{center}

\setcounter{figure}{0}
\setcounter{equation}{0}
\renewcommand\thefigure{S\arabic{figure}}
\renewcommand\thetable{S\arabic{table}}
\renewcommand\theequation{S\arabic{equation}}

\renewcommand {\thetable} {S\arabic{table}}
\renewcommand {\thefigure} {S\arabic{figure}}
\renewcommand\theequation{S\arabic{equation}}

\makeatletter
\def\@hangfrom@section#1#2#3{\@hangfrom{#1#2#3}}
\makeatother
%%%%%%%%%%%%%%%%%%%%%%%%%%%%%%%%%%%%%%%%%%%%%%%%%%%%%%
\tableofcontents

%===========================================================%
\begin{spacing}{1.5}
\section{Raman tensor tables for all MPGs}

We utilize Wolfram Mathematica to solve the symmetry-constrained Raman tensor in the main text. The character tables and symmetry operation matrices of all point groups are obtained from SpaceGroupIrep \cite{RN3438}. Since the Raman tensors depend on the choice of the coordinate system, we adopt the lattice symmetry directions of crystallographic systems, as summarized in Table \ref{Tab_LatticeDriection}. The Raman tensors for all magnetic point groups (MPGs) are listed in Table \ref{TabC1}-\ref{TabOh}. We also put the results on GitHub website [\href{https://ruichun.github.io/MagneticRamanTensor/} {https://ruichun.github.io/MagneticRamanTensor/}], where the character tables and direct product are also given. Please note that:
%%%%%%%%%%%%%%%%%%%%%%%%%%%%%%%%%%%%%%%%
\begin{table*}[htbp]
\centering
\caption{{Lattice symmetry directions of 32 original point groups.}}
\label{Tab_LatticeDriection}
\begin{threeparttable}
%\resizebox{\textwidth}{15mm}{
%\setlength{\tabcolsep}{10mm}{
\begin{tabular} {c|c|c|c|c}% {m{2.8cm}|m{3.0cm}|m{2.5cm}<{\centering}|m{2.5cm}<{\centering}|m{2.5cm}<{\centering}}
\hline
\hline
\textbf{Crystallographic system} & \textbf{Point Groups} & \textbf{Direction 1} &  \textbf{Direction 2} &  \textbf{Direction 3}  \\
\hline %%%%%%%%%%%%%%%%%%%%%%%%
Cubic	&		{$23$},
	{$m\overline{3}$},
	{$432$},
	{$\overline{4}3m$},
	{$m\overline{3}m$}
	&	$\mathbf{a}$/$\mathbf{b}$/$\mathbf{c}$	&	\begin{tabular}{l} body diagonal \\ \textit{e.g}. $\mathbf{a+b+c}$ \end{tabular}	&	\begin{tabular}{l}face diagonal  \\ \textit{e.g}. $\mathbf{a+b}$ \end{tabular}	\\ \hline
Hexagonal	&		{$6$},
	{$\overline{6}$},
	{$6/m$},
	{$622$},
	{$6mm$},
	{$\overline{6}m2$},
	{$6/mmm$}
	&	 $\mathbf{c}$	&	$\mathbf{a}$/$\mathbf{b}$	&	$\mathbf{a+2b}$/$\mathbf{2a+b}$	\\ \hline
Tetragonal	&		{$4$},
	{$\overline{4}$},
	{$4/m$},
	{$422$},
	{$4mm$},
	{$\overline{4}2m$},
	{$4/mmm$}
	&	$\mathbf{c}$	&	$\mathbf{a}$/$\mathbf{b}$	& \begin{tabular}{l}face diagonal \\ \textit{e.g.} $\mathbf{a+b}$ \end{tabular}	\\ \hline
Trigonal	&		{$3$},
	{$\overline{3}$},
	{$32$},
	{$3m$},
	{$\overline{3}m$}
	&	$\mathbf{c}$	&	$\mathbf{a}$/$\mathbf{b}$	&	NA	\\ \hline
Orthorhombic	&		{$222$},
	{$2mm$},
	{$mmm$} 	&	$\mathbf{c}$	&	$\mathbf{a}$	&	$\mathbf{b}$	\\ \hline
Monoclinic	&		{$2$},
	{$m$},
	{$2/m$}
	&	$\mathbf{c}$	&	NA	&	NA	\\ \hline
Triclinic	&		{$1$},
	{$\overline{1}$} 	&	NA	&	NA	&	NA	\\
%%%%%%%%%%%%%%%%%%%%%%%%%%%%
\hline
\hline
\end{tabular}  % }
      \begin{tablenotes}
        \footnotesize
        \item[] 1. The letters ``NA" (not available) in the third to the fifth column indicate that there are no symmetry operations along these characteristic directions.
        \item[] 2.The symbols ``/" in the third to the fifth columns denote that there are two or more equivalent character directions.
        \item[] 3. The character directions for grey MPGs and black-white MPGs obey the rules similar to those in original MPGs.
       \end{tablenotes}
\end{threeparttable}
\end{table*}
%%%%%%%%%%%%%%%%%%%%%%%%%%%%%%%%%%%%%%%%%%%%%%%%%%%%%%%%%%%%%%%%%%%%%%%

\begin{enumerate}
  \item The Raman coefficients $\alpha_{ij}(\omega)$ possess both real and imaginary parts, analogous to the dielectric function, and satisfy the Kramers-Kronig relations.

   \item  The symmetric part $\left[\alpha_{i j}^S\right]$ is  time-reversal even ($T$-even), and satisfies ${{\mathbf{\alpha }}^{S}}(-\omega )={{\left[ {{\mathbf{\alpha }}^{S}}(\omega ) \right]}^{*}}$. While the antisymmetric part $\left[\alpha_{i j}^A\right]$ is time-reversal odd ($T$-odd), and satisfies ${{\mathbf{\alpha }}^{A}}(-\omega )={{\left[ -{{\mathbf{\alpha }}^{A}}(\omega ) \right]}^{*}}$. Consequently, the $T$-even $\alpha _{ij}^{S}(0)$ may have a nonzero real component at zero frequency limit. Whereas ${{\alpha }^{A}}(0)$ vanishes, indicating  the ${{\alpha }^{A}}$ belongs a non-equilibrium tensor.

  \item The second row of each table presents the symmetric components of the Raman tensor, and the Raman tensor elements are denoted with the letter ``$\alpha$". While the remaining rows correspond to the antisymmetric components, and the Raman tensor elements are written as the letter ``A". The symmetric parts are the same as the known results in the textbooks and literature.

  \item If we assign the values +1 and $-1$ for the unitary ($R$) and anti-unitary ($R'$) operation, respectively, as the main text. In this way, the MPG $M$ can be regarded as a 1D real irrep ${{\chi}_{mag}}$ of the isomorphic nonmagnetic point group $G$, as listed in the first column of each table.
  \item For double degenerate $E$ modes and triply degenerate $T$ modes, three indices are used (e.g., $\alpha_{112}$ and $A_{212}$), where the first index denotes the number of the irreducible representation, and the last two are the Raman tensor subscripts.
  \item For the MPGs $42'2'$, $4m'm'$,  $\bar{4}2' m'$, $4/mm' m'$, $32'$, $3m'$, $\bar{6}m'2'$, $\bar{3}m'$, $62'2'$, $6m'm'$, $6/mm'm'$,  $\bar{4}'3m'$, $4'32'$, and $m\bar{3}m'$, the $E$ modes undergo splitting \cite{RN4026}, forming the chiral phonons (such as Co$_3$Sn$_2$S$_2$ \cite{RN3942}), and we indicate them by [S] after the MPG names. The spectrum characteristic of the chiral phonon in MPGs with $C_3$ symmetry is discussed in Section \ref{sec:Chiarl}.
  \item  For the origin point groups 3, 4, 6, $3/m$, $4/m$, $6/m$, $\bar{4}$, $\bar{3}$, 23, and $m\bar{3}$, the irrep $E$ modes also split, but they become degenerate again in the corresponding grey groups.
  \item For a given magnetic structure, the change of the magnetic moment direction leads to a different MPG. Therefore, compatibility relations are required to establish the correspondence between irreducible representations (irreps) before and after the change in MPG. The readers can see Section \ref{sec:NiAs} for a specific example.
  \item The Raman features under the linearly-polarized and circularly-polarized light are discussed in Section \ref{sec:AngleRaman}.

\end{enumerate}

\begin{table}[H]
\centering
\caption{The Raman tensors of the point group $1$ $(C_1)$ and corresponding MPG.}
\label{TabC1}


\end{table}
%%%%%%%%%%%%%%%%%%%%%%%%%%%%%%%%%%%%%%%%%%%%%

\begin{table}[H]
\centering
\caption{ The Raman tensors of the point group $\bar{1}$ ($C_i$) and corresponding MPGs.}
\label{TabCi}
\begin{tabular}{c|c|c}
\hline \hline
\rule{0pt}{12pt}
$\bar{1}$ ($C_i$) & ${A}_{{g}}$ & ${A}_{u}$ \\
\hline Sym. & $\left(\begin{array}{lll}\alpha_{11} & \alpha_{12} & \alpha_{13} \\ \alpha_{12} & \alpha_{22} & \alpha_{23} \\ \alpha_{13} & \alpha_{23} & \alpha_{33}\end{array}\right)$ & $\left(\begin{array}{lll}0 & 0 & 0 \\ 0 & 0 & 0 \\ 0 & 0 & 0\end{array}\right)$ \\

\hline $\bar{1}$ ($A_g$) & $\left(\begin{array}{ccc}0 & A_{12} & -A_{31} \\ -A_{12} & 0 & A_{23} \\ A_{31} & -A_{23} & 0\end{array}\right)$ & $\left(\begin{array}{lll}0 & 0 & 0 \\ 0 & 0 & 0 \\ 0 & 0 & 0\end{array}\right)$ \\

\hline $\bar{1}'$ (${A}_{u}$) & $\left(\begin{array}{lll}0 & 0 & 0 \\ 0 & 0 & 0 \\ 0 & 0 & 0\end{array}\right)$ & $\left(\begin{array}{ccc}0 & A_{12} & -A_{31} \\ -A_{12} & 0 & A_{23} \\ A_{31} & -A_{23} & 0\end{array}\right)$ \\
\hline \hline
\end{tabular}
\end{table}
%%%%%%%%%%%%%%%%%%%%%%%%%%%%%%%%%%%%%%%%%%%%%
\begin{table}[H]
\centering
\caption{ The Raman tensors of the point group 2 ($C_2$) and corresponding MPGs.}
\label{Tab1}
\begin{tabular}{c|c|c}
\hline \hline
\rule{0pt}{12pt}
2 ($C_2$) & $A$ & $B$ \\
\hline Sym. & $\left(\begin{array}{ccc}\alpha_{11} & \alpha_{12} & 0 \\ \alpha_{12} & \alpha_{22} & 0 \\ 0 & 0 & \alpha_{33}\end{array}\right)$ & $\left(\begin{array}{ccc}0 & 0 & \alpha_{13} \\ 0 & 0 & \alpha_{23} \\ \alpha_{13} & \alpha_{23} & 0\end{array}\right)$ \\

\hline 2 ($A$) & $\left(\begin{array}{ccc}0 & {A}_{12} & 0 \\ -{A}_{12} & 0 & 0 \\ 0 & 0 & 0\end{array}\right)$ & $\left(\begin{array}{ccc}0 & 0 & -{A}_{31} \\ 0 & 0 & {A}_{23} \\ {A}_{31} & -{A}_{23} & 0\end{array}\right)$ \\

\hline $2'$ ($B$) & $\left(\begin{array}{ccc}0 & 0 & -{A}_{31} \\ 0 & 0 & {A}_{23} \\ {A}_{31} & -{A}_{23} & 0\end{array}\right)$ & $\left(\begin{array}{ccc}0 & A_{12} & 0 \\ -A_{12} & 0 & 0 \\ 0 & 0 & 0\end{array}\right)$ \\
\hline \hline
\end{tabular}
\end{table}
%%%%%%%%%%%%%%%%%%%%%%%%%%%%%%%%%%%%%%%%%%%%%
\begin{table}[H]
\centering
\caption{ The Raman tensors of the point group $m$ ($C_s$) and corresponding MPGs.}
\label{TabCs}
\begin{tabular}{c|c|c}
\hline \hline
\rule{0pt}{12pt}
 $m$ ($C_s$) & ${A}^{\prime}$ & ${A}^{{\prime} {\prime}}$ \\
\hline Sym. & $\left(\begin{array}{ccc}\alpha_{11} & \alpha_{12} & 0 \\ \alpha_{12} & \alpha_{22} & 0 \\ 0 & 0 & \alpha_{33}\end{array}\right)$ & $\left(\begin{array}{ccc}0 & 0 & \alpha_{13} \\ 0 & 0 & \alpha_{23} \\ \alpha_{13} & \alpha_{23} & 0\end{array}\right)$ \\

\hline $m$ (${A}^{\prime}$) & $\left(\begin{array}{ccc}0 & {A}_{12} & 0 \\ -{A}_{12} & 0 & 0 \\ 0 & 0 & 0\end{array}\right)$ & $\left(\begin{array}{ccc}0 & 0 & -A_{31} \\ 0 & 0 & A_{23} \\ A_{31} & -A_{23} & 0\end{array}\right)$ \\

\hline ${m}^{\prime}$ (${A}''$) & $\left(\begin{array}{ccc}0 & 0 & -{A}_{31} \\ 0 & 0 & {A}_{23} \\ {A}_{31} & -{A}_{23} & 0\end{array}\right)$ & $\left(\begin{array}{ccc}0 & {A}_{12} & 0 \\ -{A}_{12} & 0 & 0 \\ 0 & 0 & 0\end{array}\right)$ \\
\hline \hline
\end{tabular}
\end{table}
%%%%%%%%%%%%%%%%%%%%%%%%%%%%%%%%%%%%%%%%%%%%%

\begin{table}[H]
\centering
\caption{ The Raman tensors of the point group $2/m$ ($C_{2h}$) and corresponding MPGs.}
\label{TabC2h}
\begin{tabular}{c|c|c|c|c}
\hline \hline
\rule{0pt}{12pt}
$2/m$ ($C_{2h}$)  & ${A}_{{g}}$ & ${A}_{u}$ & ${B}_{{g}}$ & ${B}_{u}$ \\
\hline Sym. & $\left(\begin{array}{ccc}\alpha_{11} & \alpha_{12} & 0 \\ \alpha_{12} & \alpha_{22} & 0 \\ 0 & 0 & \alpha_{33}\end{array}\right)$ & $\left(\begin{array}{lll}0 & 0 & 0 \\ 0 & 0 & 0 \\ 0 & 0 & 0\end{array}\right)$ & $\left(\begin{array}{ccc}0 & 0 & \alpha_{13} \\ 0 & 0 & \alpha_{23} \\ \alpha_{13} & \alpha_{23} & 0\end{array}\right)$ & $\left(\begin{array}{lll}0 & 0 & 0 \\ 0 & 0 & 0 \\ 0 & 0 & 0\end{array}\right)$ \\

\hline $2/m$ $({A_g})$ & $\left(\begin{array}{ccc}0 & A_{12} & 0 \\ -A_{12} & 0 & 0 \\ 0 & 0 & 0\end{array}\right)$ & $\left(\begin{array}{lll}0 & 0 & 0 \\ 0 & 0 & 0 \\ 0 & 0 & 0\end{array}\right)$ & $\left(\begin{array}{ccc}0 & 0 & -A_{31} \\ 0 & 0 & A_{23} \\ A_{31} & -A_{23} & 0\end{array}\right)$ & $\left(\begin{array}{lll}0 & 0 & 0 \\ 0 & 0 & 0 \\ 0 & 0 & 0\end{array}\right)$ \\
\hline $2/m'$ (${A}_{u}$) & $\left(\begin{array}{lll}0 & 0 & 0 \\ 0 & 0 & 0 \\ 0 & 0 & 0\end{array}\right)$ & $\left(\begin{array}{ccc}0 & A_{12} & 0 \\ -A_{12} & 0 & 0 \\ 0 & 0 & 0\end{array}\right)$ & $\left(\begin{array}{lll}0 & 0 & 0 \\ 0 & 0 & 0 \\ 0 & 0 & 0\end{array}\right)$ & $\left(\begin{array}{ccc}0 & 0 & -A_{31} \\ 0 & 0 & A_{23} \\ A_{31} & -A_{23} & 0\end{array}\right)$ \\

\hline $2^{\prime} /{m}^{\prime}$ $({B_g})$ & $\left(\begin{array}{ccc}0 & 0 & -A_{31} \\ 0 & 0 & A_{23} \\ A_{31} & -A_{23} & 0\end{array}\right)$ & $\left(\begin{array}{lll}0 & 0 & 0 \\ 0 & 0 & 0 \\ 0 & 0 & 0\end{array}\right)$ & $\left(\begin{array}{ccc}0 & A_{12} & 0 \\ -A_{12} & 0 & 0 \\ 0 & 0 & 0\end{array}\right)$ & $\left(\begin{array}{lll}0 & 0 & 0 \\ 0 & 0 & 0 \\ 0 & 0 & 0\end{array}\right)$ \\
\hline \hline
\end{tabular}
\end{table}
%%%%%%%%%%%%%%%%%%%%%%%%%%%%%%%%%%%%%%%%%%%%%
\begin{table}[H]
\centering
\caption{ The Raman tensors of the point group 222 ($D_2$) and corresponding MPGs.}
\label{TabD2}
\begin{tabular}{c|c|c|c|c}
\hline \hline
\rule{0pt}{12pt}
222 ($D_2$) & A &  $B_1$ & $B_2$ &  $B_3$ \\
\hline Sym. & $\left(\begin{array}{ccc}\alpha_{11} & 0 & 0 \\ 0 & \alpha_{22} & 0 \\ 0 & 0 & \alpha_{33}\end{array}\right)$ & $\left(\begin{array}{ccc}0 & \alpha_{12} & 0 \\ \alpha_{12} & 0 & 0 \\ 0 & 0 & 0\end{array}\right)$ & $\left(\begin{array}{ccc}0 & 0 & \alpha_{13} \\ 0 & 0 & 0 \\ \alpha_{13} & 0 & 0\end{array}\right)$ & $\left(\begin{array}{ccc}0 & 0 & 0 \\ 0 & 0 & \alpha_{23} \\ 0 & \alpha_{23} & 0\end{array}\right)$ \\

\hline 222 ($A$) & $\left(\begin{array}{lll}0 & 0 & 0 \\ 0 & 0 & 0 \\ 0 & 0 & 0\end{array}\right)$ & $\left(\begin{array}{ccc}0 & {A}_{12} & 0 \\ -{A}_{12} & 0 & 0 \\ 0 & 0 & 0\end{array}\right)$ & $\left(\begin{array}{ccc}0 & 0 & -{A}_{31} \\ 0 & 0 & 0 \\ {A}_{31} & 0 & 0\end{array}\right)$ & $\left(\begin{array}{ccc}0 & 0 & 0 \\ 0 & 0 & {A}_{23} \\ 0 & -{A}_{23} & 0\end{array}\right)$ \\

\hline $2'2'2$ (${B}_1$) & $\left(\begin{array}{ccc}0 & {A}_{12} & 0 \\ -{A}_{12} & 0 & 0 \\ 0 & 0 & 0\end{array}\right)$ & $\left(\begin{array}{lll}0 & 0 & 0 \\ 0 & 0 & 0 \\ 0 & 0 & 0\end{array}\right)$ & $\left(\begin{array}{ccc}0 & 0 & 0 \\ 0 & 0 & {A}_{23} \\ 0 & -{A}_{23} & 0\end{array}\right)$ & $\left(\begin{array}{ccc}0 & 0 & -{A}_{31} \\ 0 & 0 & 0 \\ {A}_{31} & 0 & 0\end{array}\right)$ \\

\hline $2'22'$ (${B}_2$) & $\left(\begin{array}{ccc}0 & 0 & -{A}_{31} \\ 0 & 0 & 0 \\ {A}_{31} & 0 & 0\end{array}\right)$ & $\left(\begin{array}{ccc}0 & 0 & 0 \\ 0 & 0 & {A}_{23} \\ 0 & -{A}_{23} & 0\end{array}\right)$ & $\left(\begin{array}{lll}0 & 0 & 0 \\ 0 & 0 & 0 \\ 0 & 0 & 0\end{array}\right)$ & $\left(\begin{array}{ccc}0 & {A}_{12} & 0 \\ -{A}_{12} & 0 & 0 \\ 0 & 0 & 0\end{array}\right)$ \\
\hline \hline
\end{tabular}
\end{table}
%%%%%%%%%%%%%%%%%%%%%%%%%%%%%%%%%%%%%%%%%%%%%

\begin{table}[H]
\centering
\caption{ The Raman tensors of the point group $2mm$ ($C_{2v}$) and corresponding MPGs.}
\label{TabC2v}
\begin{tabular}{c|c|c|c|c}
\hline  \hline
\rule{0pt}{12pt}
$2mm$ ($C_{2v}$) & ${A}_1$ & ${A}_2$ & ${B}_1$ &  $B_2$ \\
\hline Sym. & $\left(\begin{array}{ccc}\alpha_{11} & 0 & 0 \\ 0 & \alpha_{22} & 0 \\ 0 & 0 & \alpha_{33}\end{array}\right)$ & $\left(\begin{array}{ccc}0 & \alpha_{12} & 0 \\ \alpha_{12} & 0 & 0 \\ 0 & 0 & 0\end{array}\right)$ & $\left(\begin{array}{ccc}0 & 0 & \alpha_{13} \\ 0 & 0 & 0 \\ \alpha_{13} & 0 & 0\end{array}\right)$ & $\left(\begin{array}{ccc}0 & 0 & 0 \\ 0 & 0 & \alpha_{23} \\ 0 & \alpha_{23} & 0\end{array}\right)$ \\

\hline $mm2$ (${A}_1$) & $\left(\begin{array}{lll}0 & 0 & 0 \\ 0 & 0 & 0 \\ 0 & 0 & 0\end{array}\right)$ & $\left(\begin{array}{ccc}0 & {A}_{12} & 0 \\ -{A}_{12} & 0 & 0 \\ 0 & 0 & 0\end{array}\right)$ & $\left(\begin{array}{ccc}0 & 0 & -{A}_{31} \\ 0 & 0 & 0 \\ {A}_{31} & 0 & 0\end{array}\right)$ & $\left(\begin{array}{ccc}0 & 0 & 0 \\ 0 & 0 & {A}_{23} \\ 0 & -{A}_{23} & 0\end{array}\right)$ \\

\hline $m'm'2$ (${A}_2$) & $\left(\begin{array}{ccc}0 & {A}_{12} & 0 \\ -{A}_{12} & 0 & 0 \\ 0 & 0 & 0\end{array}\right)$ & $\left(\begin{array}{lll}0 & 0 & 0 \\ 0 & 0 & 0 \\ 0 & 0 & 0\end{array}\right)$ & $\left(\begin{array}{ccc}0 & 0 & 0 \\ 0 & 0 & {A}_{23} \\ 0 & -{A}_{23} & 0\end{array}\right)$ & $\left(\begin{array}{ccc}0 & 0 & -{A}_{31} \\ 0 & 0 & 0 \\ {A}_{31} & 0 & 0\end{array}\right)$ \\

\hline $m'm2'$ (${B}_1$) & $\left(\begin{array}{ccc}0 & 0 & -A_{31} \\ 0 & 0 & 0 \\ A_{31} & 0 & 0\end{array}\right)$ & $\left(\begin{array}{ccc}0 & 0 & 0 \\ 0 & 0 & {A}_{23} \\ 0 & -{A}_{23} & 0\end{array}\right)$ & $\left(\begin{array}{lll}0 & 0 & 0 \\ 0 & 0 & 0 \\ 0 & 0 & 0\end{array}\right)$ & $\left(\begin{array}{ccc}0 & {A}_{12} & 0 \\ -{A}_{12} & 0 & 0 \\ 0 & 0 & 0\end{array}\right)$ \\
\hline \hline
\end{tabular}
\end{table}
%%%%%%%%%%%%%%%%%%%%%%%%%%%%%%%%%%%%%%%%%%%%%
\begin{table}[H]
\centering
\caption{ The Raman tensors of the point group $mmm$ ($D_{2h}$) and corresponding MPGs.}
\label{TabD2h}
\resizebox{\textwidth}{!}{%
\begin{tabular}{c|c|c|c|c|c|c|c|c}
\hline \hline
\rule{0pt}{12pt}
$mmm$ ($D_{2h}$) & ${A}_{{g}}$ & ${A}_{u}$ & ${B}_{1 g}$ & ${B}_{1 u}$ & ${B}_{2 g}$ & $B_{2u}$ & ${B}_{{3 g}}$ & ${B}_{3 u}$ \\
\hline Sym. & $\left(\begin{array}{ccc}\alpha_{11} & 0 & 0 \\ 0 & \alpha_{22} & 0 \\ 0 & 0 & \alpha_{33}\end{array}\right)$ & $\left(\begin{array}{lll}0 & 0 & 0 \\ 0 & 0 & 0 \\ 0 & 0 & 0\end{array}\right)$ & $\left(\begin{array}{ccc}0 & \alpha_{12} & 0 \\ \alpha_{12} & 0 & 0 \\ 0 & 0 & 0\end{array}\right)$ & $\left(\begin{array}{lll}0 & 0 & 0 \\ 0 & 0 & 0 \\ 0 & 0 & 0\end{array}\right)$ & $\left(\begin{array}{ccc}0 & 0 & \alpha_{13} \\ 0 & 0 & 0 \\ \alpha_{13} & 0 & 0\end{array}\right)$ & $\left(\begin{array}{lll}0 & 0 & 0 \\ 0 & 0 & 0 \\ 0 & 0 & 0\end{array}\right)$ & $\left(\begin{array}{ccc}0 & 0 & 0 \\ 0 & 0 & \alpha_{23} \\ 0 & \alpha_{23} & 0\end{array}\right)$ & $\left(\begin{array}{lll}0 & 0 & 0 \\ 0 & 0 & 0 \\ 0 & 0 & 0\end{array}\right)$ \\

\hline $mmm$ ($A_g$) & $\left(\begin{array}{lll}0 & 0 & 0 \\ 0 & 0 & 0 \\ 0 & 0 & 0\end{array}\right)$ & $\left(\begin{array}{lll}0 & 0 & 0 \\ 0 & 0 & 0 \\ 0 & 0 & 0\end{array}\right)$ & $\left(\begin{array}{ccc}0 & {A}_{12} & 0 \\ -{A}_{12} & 0 & 0 \\ 0 & 0 & 0\end{array}\right)$ & $\left(\begin{array}{lll}0 & 0 & 0 \\ 0 & 0 & 0 \\ 0 & 0 & 0\end{array}\right)$ & $\left(\begin{array}{ccc}0 & 0 & -{A}_{31} \\ 0 & 0 & 0 \\ {A}_{31} & 0 & 0\end{array}\right)$ & $\left(\begin{array}{lll}0 & 0 & 0 \\ 0 & 0 & 0 \\ 0 & 0 & 0\end{array}\right)$ & $\left(\begin{array}{ccc}0 & 0 & 0 \\ 0 & 0 & {A}_{23} \\ 0 & -{A}_{23} & 0\end{array}\right)$ & $\left(\begin{array}{lll}0 & 0 & 0 \\ 0 & 0 & 0 \\ 0 & 0 & 0\end{array}\right)$ \\
\hline \hline
\end{tabular}
}
\end{table}
%%%%%%%%%%%%%%%%%%%%%%%%%%%%%%%%%%%%%%%%%%%%%

\begin{table}[H]
\centering
\caption{ The Raman tensors of the point group 4 ($C_4$) and corresponding MPGs.}
\label{TabC4}
%
\end{table}
%%%%%%%%%%%%%%%%%%%%%%%%%%%%%%%%%%%%%%%%%%%%%
\begin{table}[H]
\centering
\caption{ The Raman tensors of the point group $\bar{4}$ ($S_4$) and corresponding MPGs.}
\label{TabS4}
%
\end{table}
%%%%%%%%%%%%%%%%%%%%%%%%%%%%%%%%%%%%%%%%%%%%%
\begin{table}[H]
\centering
\caption{ The Raman tensors of the point group $4/m$ ($C_{4h}$) and corresponding MPGs.}
\label{TabC4h}
\resizebox{\textwidth}{!}{%
\begin{tabular}{c|c|c|c|c|c|c|c|c}
\hline \hline
\rule{0pt}{12pt}
$4/m$ ($C_{4h}$) & ${A}_{{g}}$ & ${A}_u$ & ${B}_{{g}}$ & ${B}_u$ & ${ }^1 {E}_{{g}}$ & ${ }^2 {E}_{{g}}$ & ${ }^1 {E}_u$ & ${ }^2 {E}_u$ \\
\hline Sym. & $\left(\begin{array}{ccc}\alpha_{11} & 0 & 0 \\ 0 & \alpha_{11} & 0 \\ 0 & 0 & \alpha_{33}\end{array}\right)$ & $\left(\begin{array}{lll}0 & 0 & 0 \\ 0 & 0 & 0 \\ 0 & 0 & 0\end{array}\right)$ & $\left(\begin{array}{ccc}\alpha_{11} & \alpha_{12} & 0 \\ \alpha_{12} & -\alpha_{11} & 0 \\ 0 & 0 & 0\end{array}\right)$ & $\left(\begin{array}{lll}0 & 0 & 0 \\ 0 & 0 & 0 \\ 0 & 0 & 0\end{array}\right)$ & $\left(\begin{array}{ccc}0 & 0 & \alpha_{13} \\ 0 & 0 & -i \alpha_{13} \\ \alpha_{13} & -i \alpha_{13} & 0\end{array}\right)$ & $\left(\begin{array}{ccc}0 & 0 & \alpha_{13} \\ 0 & 0 & i \alpha_{13} \\ \alpha_{13} & i \alpha_{13} & 0\end{array}\right)$ & $\left(\begin{array}{lll}0 & 0 & 0 \\ 0 & 0 & 0 \\ 0 & 0 & 0\end{array}\right)$ & $\left(\begin{array}{lll}0 & 0 & 0 \\ 0 & 0 & 0 \\ 0 & 0 & 0\end{array}\right)$ \\

\hline $4/m$ ($A_g$) & $\left(\begin{array}{ccc}0 & A_{12} & 0 \\ -A_{12} & 0 & 0 \\ 0 & 0 & 0\end{array}\right)$ & $\left(\begin{array}{lll}0 & 0 & 0 \\ 0 & 0 & 0 \\ 0 & 0 & 0\end{array}\right)$ & $\left(\begin{array}{lll}0 & 0 & 0 \\ 0 & 0 & 0 \\ 0 & 0 & 0\end{array}\right)$ & $\left(\begin{array}{lll}0 & 0 & 0 \\ 0 & 0 & 0 \\ 0 & 0 & 0\end{array}\right)$ & $\left(\begin{array}{ccc}0 & 0 & -{A}_{31} \\ 0 & 0 & i {A}_{31} \\ {A}_{31} & -i {A}_{31} & 0\end{array}\right)$ & $\left(\begin{array}{ccc}0 & 0 & -{A}_{31} \\ 0 & 0 & -i {A}_{31} \\ {A}_{31} & i {A}_{31} & 0\end{array}\right)$ & $\left(\begin{array}{lll}0 & 0 & 0 \\ 0 & 0 & 0 \\ 0 & 0 & 0\end{array}\right)$ & $\left(\begin{array}{lll}0 & 0 & 0 \\ 0 & 0 & 0 \\ 0 & 0 & 0\end{array}\right)$ \\
\hline \hline
\end{tabular}
}
\end{table}
%%%%%%%%%%%%%%%%%%%%%%%%%%%%%%%%%%%%%%%%%%%%%

\begin{table}[H]
\centering
\caption{ The Raman tensors of the point group 422 ($D_4$) and corresponding MPGs.}
\label{TabD4}
\begin{tabular}{c|c|c|c|c|c}
\hline \hline
\rule{0pt}{12pt}
422 ($D_4$) & ${A}_1$ & ${A}_2$ & ${B}_{{1}}$ & ${B}_2$ & $E$ \\
\hline Sym. & $\left(\begin{array}{ccc}\alpha_{11} & 0 & 0 \\ 0 & \alpha_{11} & 0 \\ 0 & 0 & \alpha_{33}\end{array}\right)$ & $\left(\begin{array}{lll}0 & 0 & 0 \\ 0 & 0 & 0 \\ 0 & 0 & 0\end{array}\right)$ & $\left(\begin{array}{ccc}\alpha_{11} & 0 & 0 \\ 0 & -\alpha_{11} & 0 \\ 0 & 0 & 0\end{array}\right)$ & $\left(\begin{array}{ccc}0 & \alpha_{12} & 0 \\ \alpha_{12} & 0 & 0 \\ 0 & 0 & 0\end{array}\right)$ & $
\begin{aligned}
& \left(\begin{array}{ccc}
0 & 0 & 0 \\
0 & 0 & \alpha_{123} \\
0 & \alpha_{123} & 0
\end{array}\right) \\
& \left(\begin{array}{ccc}
0 & 0 & \alpha_{123} \\
0 & 0 & 0 \\
\alpha_{123} & 0 & 0
\end{array}\right)
\end{aligned}
$ \\

\hline 422 (${A}_1$) & $\left(\begin{array}{lll}0 & 0 & 0 \\ 0 & 0 & 0 \\ 0 & 0 & 0\end{array}\right)$ & $\left(\begin{array}{ccc}0 & {A}_{12} & 0 \\ -{A}_{12} & 0 & 0 \\ 0 & 0 & 0\end{array}\right)$ & $\left(\begin{array}{lll}0 & 0 & 0 \\ 0 & 0 & 0 \\ 0 & 0 & 0\end{array}\right)$ & $\left(\begin{array}{lll}0 & 0 & 0 \\ 0 & 0 & 0 \\ 0 & 0 & 0\end{array}\right)$ &
$
\begin{aligned}
& \left(\begin{array}{ccc}
0 & 0 & 0 \\
0 & 0 & {A}_{123} \\
0 & -{A}_{123} & 0
\end{array}\right) \\
& \left(\begin{array}{ccc}
0 & 0 & {A}_{123} \\
0 & 0 & 0 \\
-{A}_{123} & 0 & 0
\end{array}\right)
\end{aligned}
$ \\

\hline $42'2'$ (${A}_2$) [S]& $\left(\begin{array}{ccc}0 & {A}_{12} & 0 \\ -{A}_{12} & 0 & 0 \\ 0 & 0 & 0\end{array}\right)$ & $\left(\begin{array}{lll}0 & 0 & 0 \\ 0 & 0 & 0 \\ 0 & 0 & 0\end{array}\right)$ & $\left(\begin{array}{lll}0 & 0 & 0 \\ 0 & 0 & 0 \\ 0 & 0 & 0\end{array}\right)$ & $\left(\begin{array}{lll}0 & 0 & 0 \\ 0 & 0 & 0 \\ 0 & 0 & 0\end{array}\right)$ & $
\begin{aligned}
& \left(\begin{array}{ccc}
0 & 0 & {A}_{113} \\
0 & 0 & 0 \\
-{A}_{113} & 0 & 0
\end{array}\right) \\
& \left(\begin{array}{ccc}
0 & 0 & 0 \\
0 & 0 & -{A}_{113} \\
0 & {A}_{113} & 0
\end{array}\right)
\end{aligned}
$ \\
\hline \hline
\end{tabular}
\end{table}
%%%%%%%%%%%%%%%%%%%%%%%%%%%%%%%%%%%%%%%%%%%%%
\begin{table}[H]
\centering
\caption{ The Raman tensors of the point group $4mm$ ($C_{4v}$) and corresponding MPGs.}
\label{TabC4v}
\begin{tabular}{c|c|c|c|c|c}
\hline \hline
\rule{0pt}{12pt}
$4mm$ ($C_{4v}$) & ${A}_1$ & ${A}_{{2}}$ & ${B}_1$ &  $B_2$ & E \\
\hline Sym. & $\left(\begin{array}{ccc}\alpha_{11} & 0 & 0 \\ 0 & \alpha_{11} & 0 \\ 0 & 0 & \alpha_{33}\end{array}\right)$ & $\left(\begin{array}{lll}0 & 0 & 0 \\ 0 & 0 & 0 \\ 0 & 0 & 0\end{array}\right)$ & $\left(\begin{array}{ccc}\alpha_{11} & 0 & 0 \\ 0 & -\alpha_{11} & 0 \\ 0 & 0 & 0\end{array}\right)$ & $\left(\begin{array}{ccc}0 & \alpha_{12} & 0 \\ \alpha_{12} & 0 & 0 \\ 0 & 0 & 0\end{array}\right)$ & $
\begin{gathered}
\left(\begin{array}{ccc}
0 & 0 & \alpha_{113} \\
0 & 0 & 0 \\
\alpha_{113} & 0 & 0
\end{array}\right) \\
\left(\begin{array}{ccc}
0 & 0 & 0 \\
0 & 0 & -\alpha_{113} \\
0 & -\alpha_{113} & 0
\end{array}\right)
\end{gathered}
$ \\

\hline $4mm$ (${A}_1$) & $\left(\begin{array}{lll}0 & 0 & 0 \\ 0 & 0 & 0 \\ 0 & 0 & 0\end{array}\right)$ & $\left(\begin{array}{ccc}0 & {A}_{12} & 0 \\ -{A}_{12} & 0 & 0 \\ 0 & 0 & 0\end{array}\right)$ & $\left(\begin{array}{lll}0 & 0 & 0 \\ 0 & 0 & 0 \\ 0 & 0 & 0\end{array}\right)$ & $\left(\begin{array}{lll}0 & 0 & 0 \\ 0 & 0 & 0 \\ 0 & 0 & 0\end{array}\right)$ & $
\begin{aligned}
& \left(\begin{array}{ccc}
0 & 0 & {A}_{113} \\
0 & 0 & 0 \\
-{A}_{113} & 0 & 0
\end{array}\right) \\
& \left(\begin{array}{ccc}
0 & 0 & 0 \\
0 & 0 & -{A}_{113} \\
0 & {A}_{113} & 0
\end{array}\right)
\end{aligned}
$ \\
\hline \hline
\end{tabular}
\end{table}
%%%%%%%%%%%%%%%%%%%%%%%%%%%%%%%%%%%%%%%%%%%%%
\begin{table}[H]
\centering
\caption{ The Raman tensors of the point group $\bar{4}2m$ ($D_{2d}$) and corresponding MPGs.}
\label{TabD2d}
\resizebox{\textwidth}{!}{%
\begin{tabular}{c|c|c|c|c|c}
\hline \hline
\rule{0pt}{12pt}
$\bar{4}2m$ ($D_{2d}$) & ${A}_1$ & ${A}_2$ & ${B}_{{1}}$ & ${B}_{{2}}$ & $E$ \\
\hline
Sym. & $\left(\begin{array}{ccc}\alpha_{11} & 0 & 0 \\ 0 & \alpha_{11} & 0 \\ 0 & 0 & \alpha_{33}\end{array}\right)$ & $\left(\begin{array}{lll}0 & 0 & 0 \\ 0 & 0 & 0 \\ 0 & 0 & 0\end{array}\right)$ & $\left(\begin{array}{ccc}\alpha_{11} & 0 & 0 \\ 0 & -\alpha_{11} & 0 \\ 0 & 0 & 0\end{array}\right)$ & $\left(\begin{array}{ccc}0 & \alpha_{12} & 0 \\ \alpha_{12} & 0 & 0 \\ 0 & 0 & 0\end{array}\right)$ & $
\begin{gathered}
\left(\begin{array}{ccc}
0 & 0 & 0 \\
0 & 0 & \alpha_{123} \\
0 & \alpha_{123} & 0
\end{array}\right) \\
\left(\begin{array}{ccc}
0 & 0 & -\alpha_{123} \\
0 & 0 & 0 \\
-\alpha_{123} & 0 & 0
\end{array}\right)
\end{gathered}
$ \\

\hline $\bar{4}2m$ (${A}_1$) & $\left(\begin{array}{lll}0 & 0 & 0 \\ 0 & 0 & 0 \\ 0 & 0 & 0\end{array}\right)$ & $\left(\begin{array}{ccc}0 & {A}_{12} & 0 \\ -{A}_{12} & 0 & 0 \\ 0 & 0 & 0\end{array}\right)$ & $\left(\begin{array}{lll}0 & 0 & 0 \\ 0 & 0 & 0 \\ 0 & 0 & 0\end{array}\right)$ & $\left(\begin{array}{lll}0 & 0 & 0 \\ 0 & 0 & 0 \\ 0 & 0 & 0\end{array}\right)$ & $
\begin{aligned}
& \left(\begin{array}{ccc}
0 & 0 & 0 \\
0 & 0 & {A}_{123} \\
0 & -{A}_{123} & 0
\end{array}\right) \\
& \left(\begin{array}{ccc}
0 & 0 & -{A}_{123} \\
0 & 0 & 0 \\
{A}_{123} & 0 & 0
\end{array}\right)
\end{aligned}
$ \\
\hline \hline
\end{tabular}
}
\end{table}
%%%%%%%%%%%%%%%%%%%%%%%%%%%%%%%%%%%%%%%%%%%%%
\begin{table}[H]
\centering
\caption{ The Raman tensors of the point group $4/mmm$ ($D_{4h}$) and corresponding MPGs.}
\label{TabD4h}
\resizebox{\textwidth}{!}{%
\begin{tabular}{c|c|c|c|c|c|c|c|c|c|c}
\hline \hline
\rule{0pt}{12pt}
$4/mmm$ ($D_{4h}$) & ${A}_{{1  g}}$ & ${A}_{1 {u}}$ & ${A}_{{2  g}}$ & ${A}_{{2  u}}$ & ${B}_{{1  g}}$ & ${B}_{1 {u}}$ & ${B}_{{2  g}}$ &  $B_{2  {u}}$ & ${E}_{{g}}$ & ${E}_{{u}}$ \\
\hline Sym. & $\left(\begin{array}{ccc}\alpha_{11} & 0 & 0 \\ 0 & \alpha_{11} & 0 \\ 0 & 0 & \alpha_{33}\end{array}\right)$ & $\left(\begin{array}{lll}0 & 0 & 0 \\ 0 & 0 & 0 \\ 0 & 0 & 0\end{array}\right)$ & $\left(\begin{array}{lll}0 & 0 & 0 \\ 0 & 0 & 0 \\ 0 & 0 & 0\end{array}\right)$ & $\left(\begin{array}{lll}0 & 0 & 0 \\ 0 & 0 & 0 \\ 0 & 0 & 0\end{array}\right)$ & $\left(\begin{array}{ccc}\alpha_{11} & 0 & 0 \\ 0 & -\alpha_{11} & 0 \\ 0 & 0 & 0\end{array}\right)$ & $\left(\begin{array}{lll}0 & 0 & 0 \\ 0 & 0 & 0 \\ 0 & 0 & 0\end{array}\right)$ & $\left(\begin{array}{ccc}0 & \alpha_{12} & 0 \\ \alpha_{12} & 0 & 0 \\ 0 & 0 & 0\end{array}\right)$ & $\left(\begin{array}{lll}0 & 0 & 0 \\ 0 & 0 & 0 \\ 0 & 0 & 0\end{array}\right)$ &

$\begin{aligned}
& \left(\begin{array}{ccc}
0 & 0 & 0 \\
0 & 0 & \alpha_{123} \\
0 & \alpha_{123} & 0
\end{array}\right)
\\&
\left(\begin{array}{ccc}
0 & 0 & \alpha_{123} \\
0 & 0 & 0 \\
\alpha_{123} & 0 & 0
\end{array}\right)
\end{aligned}$

&
%%%%xrcxrc
$\begin{aligned}
 & \left(\begin{array}{lll}0 & 0 & 0 \\ 0 & 0 & 0 \\ 0 & 0 & 0\end{array}\right) \\&
 \left(\begin{array}{lll}0 & 0 & 0 \\ 0 & 0 & 0 \\ 0 & 0 & 0\end{array}\right)
 \end{aligned}$
 \\
\hline
\hline
\end{tabular}
}
\end{table}
%%%%%%%%%%%%%%%%%%%%%%%%%%%%%%%%%%%%%%%%%%%%%%%%%
\begin{table}[H]
\centering
\caption{ The Raman tensors of the point group 3 ($C_3$) and corresponding MPGs.}
\label{TabC3}
\begin{tabular}{c|c|c|c}
\hline \hline
\rule{0pt}{12pt}
3 ($C_3$) & $A$ & ${ }^1 E$ & ${ }^2 E$ \\
\hline Sym. & $\left(\begin{array}{ccc}\alpha_{11} & 0 & 0 \\ 0 & \alpha_{11} & 0 \\ 0 & 0 & \alpha_{33}\end{array}\right)$ & $\left(\begin{array}{ccc}\alpha_{11} & i \alpha_{11} & \alpha_{13} \\ i \alpha_{11} & -\alpha_{11} & -i \alpha_{13} \\ \alpha_{13} & -i \alpha_{13} & 0\end{array}\right)$ & $\left(\begin{array}{ccc}\alpha_{11} & -i \alpha_{11} & \alpha_{13} \\ -i \alpha_{11} & -\alpha_{11} & i \alpha_{13} \\ \alpha_{13} & i \alpha_{13} & 0\end{array}\right)$ \\
\hline 3 ($A$) & $\left(\begin{array}{ccc}0 & {A}_{12} & 0 \\ -{A}_{12} & 0 & 0 \\ 0 & 0 & 0\end{array}\right)$ & $\left(\begin{array}{ccc}0 & 0 & -{A}_{31} \\ 0 & 0 & i {A}_{31} \\ {A}_{31} & -i {A}_{31} & 0\end{array}\right)$ & $\left(\begin{array}{ccc}0 & 0 & -{A}_{31} \\ 0 & 0 & -i {A}_{31} \\ {A}_{31} & i {A}_{31} & 0\end{array}\right)$ \\
\hline \hline
\end{tabular}
\end{table}
%%%%%%%%%%%%%%%%%%%%%%%%%%%%%%%%%%%%%%%%%%%%%

\begin{table}[H]
\centering
\caption{ The Raman tensors of the point group $\bar{3}$ ($S_6$) and corresponding MPGs.}
\label{TabS6}
\resizebox{\textwidth}{!}{%
\begin{tabular}{c|c|c|c|c|c|c}
\hline \hline
\rule{0pt}{12pt}
$\bar{3}$ ($S_6$) & ${A}_{{g}}$ & ${A}_u$ & ${ }^1 {E}_{{g}}$ & ${ }^2 {E}_{{g}}$ & ${ }^1 {E}_u$ & ${ }^2 {E}_u$ \\
\hline Sym. & $\left(\begin{array}{ccc}\alpha_{11} & 0 & 0 \\ 0 & \alpha_{11} & 0 \\ 0 & 0 & \alpha_{33}\end{array}\right)$ & $\left(\begin{array}{lll}0 & 0 & 0 \\ 0 & 0 & 0 \\ 0 & 0 & 0\end{array}\right)$ & $\left(\begin{array}{ccc}\alpha_{11} & i \alpha_{11} & \alpha_{13} \\ i \alpha_{11} & -\alpha_{11} & -i \alpha_{13} \\ \alpha_{13} & -i \alpha_{13} & 0\end{array}\right)$ & $\left(\begin{array}{ccc}\alpha_{11} & -i \alpha_{11} & \alpha_{13} \\ -i \alpha_{11} & -\alpha_{11} & i \alpha_{13} \\ \alpha_{13} & i \alpha_{13} & 0\end{array}\right)$ & $\left(\begin{array}{lll}0 & 0 & 0 \\ 0 & 0 & 0 \\ 0 & 0 & 0\end{array}\right)$ & $\left(\begin{array}{lll}0 & 0 & 0 \\ 0 & 0 & 0 \\ 0 & 0 & 0\end{array}\right)$ \\

\hline $\bar{3}$ (${A}_{{g}}$) & $\left(\begin{array}{ccc}0 & {A}_{12} & 0 \\ -{A}_{12} & 0 & 0 \\ 0 & 0 & 0\end{array}\right)$ & $\left(\begin{array}{lll}0 & 0 & 0 \\ 0 & 0 & 0 \\ 0 & 0 & 0\end{array}\right)$ & $\left(\begin{array}{ccc}0 & 0 & -{A}_{31} \\ 0 & 0 & i {A}_{31} \\ {A}_{31} & -i {A}_{31} & 0\end{array}\right)$ & $\left(\begin{array}{ccc}0 & 0 & -{A}_{31} \\ 0 & 0 & -i {A}_{31} \\ {A}_{31} & i {A}_{31} & 0\end{array}\right)$ & $\left(\begin{array}{lll}0 & 0 & 0 \\ 0 & 0 & 0 \\ 0 & 0 & 0\end{array}\right)$ & $\left(\begin{array}{lll}0 & 0 & 0 \\ 0 & 0 & 0 \\ 0 & 0 & 0\end{array}\right)$ \\
\hline \hline
\end{tabular}
}
\end{table}
%%%%%%%%%%%%%%%%%%%%%%%%%%%%%%

%%%%%%%%%%%%%%%%%%%%%%%%%%%%%%%%%%%%%%%%%%%%%
\begin{table}[H]
\centering
\caption{ The Raman tensors of the point group 32 ($D_3$) and corresponding MPGs.}
\label{TabD3}
\begin{tabular}{c|c|c|c}
\hline \hline
\rule{0pt}{12pt}
32 ($D_3$) & ${A}_1$ & ${A}_2$ & E \\
\hline Sym. & $\left(\begin{array}{ccc}\alpha_{11} & 0 & 0 \\ 0 & \alpha_{11} & 0 \\ 0 & 0 & \alpha_{33}\end{array}\right)$ & $\left(\begin{array}{lll}0 & 0 & 0 \\ 0 & 0 & 0 \\ 0 & 0 & 0\end{array}\right)$ & $\left(\begin{array}{ccc}\alpha_{111} & 0 & \alpha_{113} \\ 0 & -\alpha_{111} & 0 \\ \alpha_{113} & 0 & 0\end{array}\right)$
$\left(\begin{array}{ccc}0 & -\alpha_{111} & 0 \\ -\alpha_{111} & 0 & \alpha_{113} \\ 0 & \alpha_{113} & 0\end{array}\right)$ \\

\hline 32 (${A}_1$) & $\left(\begin{array}{lll}0 & 0 & 0 \\ 0 & 0 & 0 \\ 0 & 0 & 0\end{array}\right)$ & $\left(\begin{array}{ccc}0 & A_{12} & 0 \\ -A_{12} & 0 & 0 \\ 0 & 0 & 0\end{array}\right)$ & $\left(\begin{array}{ccc}0 & 0 & A_{113} \\ 0 & 0 & 0 \\ -A_{113} & 0 & 0\end{array}\right)$
$\left(\begin{array}{ccc}0 & 0 & 0 \\ 0 & 0 & A_{113} \\ 0 & -A_{113} & 0\end{array}\right)$ \\

\hline $32'$ ($A_2$) [S] & $\left(\begin{array}{ccc}0 & {A}_{12} & 0 \\ -{A}_{12} & 0 & 0 \\ 0 & 0 & 0\end{array}\right)$ & $\left(\begin{array}{lll}0 & 0 & 0 \\ 0 & 0 & 0 \\ 0 & 0 & 0\end{array}\right)$ & $\left(\begin{array}{ccc}0 & 0 & 0 \\ 0 & 0 & A_{123} \\ 0 & -A_{123} & 0\end{array}\right)$
$\left(\begin{array}{ccc}0 & 0 & -A_{123} \\ 0 & 0 & 0 \\ A_{123} & 0 & 0\end{array}\right)$ \\
\hline \hline
\end{tabular}
\end{table}
%%%%%%%%%%%%%%%%%%%%%%%%%%%%%%%%%%%%%%%%%%%%%

\begin{table}
\centering
\caption{ The Raman tensors of the point group $3m$ ($C_{3v}$) and corresponding MPGs.}
\label{TabC3v}
\begin{tabular}{c|c|c|c}
\hline \hline
\rule{0pt}{12pt}
 $3m$ ($C_{3v}$) & ${A}_1$ & ${A}_2$ & E \\
\hline Sym. & $\left(\begin{array}{ccc}\alpha_{11} & 0 & 0 \\ 0 & \alpha_{11} & 0 \\ 0 & 0 & \alpha_{33}\end{array}\right)$ & $\left(\begin{array}{lll}0 & 0 & 0 \\ 0 & 0 & 0 \\ 0 & 0 & 0\end{array}\right)$ & $
\begin{gathered}
\left(\begin{array}{ccc}
\alpha_{111} & 0 & 0 \\
0 & -\alpha_{111} & \alpha_{123} \\
0 & \alpha_{123} & 0
\end{array}\right) \\
\left(\begin{array}{ccc}
0 & -\alpha_{111} & -\alpha_{123} \\
-\alpha_{111} & 0 & 0 \\
-\alpha_{123} & 0 & 0
\end{array}\right)
\end{gathered}
$ \\

\hline $3m$ (${A}_1$) & $\left(\begin{array}{lll}0 & 0 & 0 \\ 0 & 0 & 0 \\ 0 & 0 & 0\end{array}\right)$ & $\left(\begin{array}{ccc}0 & {A}_{12} & 0 \\ -{A}_{12} & 0 & 0 \\ 0 & 0 & 0\end{array}\right)$ & $
\begin{aligned}
& \left(\begin{array}{ccc}
0 & 0 & 0 \\
0 & 0 & {A}_{123} \\
0 & -{A}_{123} & 0
\end{array}\right) \\
& \left(\begin{array}{ccc}
0 & 0 & -{A}_{123} \\
0 & 0 & 0 \\
{A}_{123} & 0 & 0
\end{array}\right)
\end{aligned}
$ \\

\hline $3m^{\prime} $ ($A_2$) [S] & $\left(\begin{array}{ccc}0 & A_{12} & 0 \\ -A_{12} & 0 & 0 \\ 0 & 0 & 0\end{array}\right)$ & $\left(\begin{array}{lll}0 & 0 & 0 \\ 0 & 0 & 0 \\ 0 & 0 & 0\end{array}\right)$ & $\left(\begin{array}{ccc}0 & 0 & A_{113} \\ 0 & 0 & 0 \\ -A_{113} & 0 & 0\end{array}\right)$
$\left(\begin{array}{ccc}0 & 0 & 0 \\ 0 & 0 & A_{113} \\ 0 & -A_{113} & 0\end{array}\right)$ \\
\hline \hline
\end{tabular}
\end{table}

%%%%%%%%%%%%%%%%%%%%%%%%%%%%%%%%%%%%%%%%%%%%%

\begin{table}
\centering
\caption{ The Raman tensors of the point group $\bar{3}{m}$ ($D_{3d}$) and corresponding MPGs.}
\label{TabD3d}

\resizebox{\textwidth}{!}{%
%%%%%xrc
\begin{tabular}{c|c|c|c|c|c|c}
\hline \hline
\rule{0pt}{12pt}
$\bar{3}{m}$ ($D_{3d}$) & ${A}_{{1  g}}$ & ${A}_{{1u}}$ & ${A}_{{2g}}$ & ${A}_{{2u}}$ & ${E}_{{g}}$ & ${E}_u$ \\
\hline Sym. & $\left(\begin{array}{ccc}\alpha_{11} & 0 & 0 \\ 0 & \alpha_{11} & 0 \\ 0 & 0 & \alpha_{33}\end{array}\right)$ & $\left(\begin{array}{lll}0 & 0 & 0 \\ 0 & 0 & 0 \\ 0 & 0 & 0\end{array}\right)$ & $\left(\begin{array}{lll}0 & 0 & 0 \\ 0 & 0 & 0 \\ 0 & 0 & 0\end{array}\right)$ & $\left(\begin{array}{lll}0 & 0 & 0 \\ 0 & 0 & 0 \\ 0 & 0 & 0\end{array}\right)$
&
$\begin{aligned} &
\left(\begin{array}{ccc}\alpha_{111} & 0 & \alpha_{113} \\ 0 & -\alpha_{111} & 0 \\ \alpha_{113} & 0 & 0\end{array}\right)
\\&
\left(\begin{array}{ccc}0 & \alpha_{111} & 0 \\ \alpha_{111} & 0 & -\alpha_{113} \\ 0 & -\alpha_{113} & 0\end{array}\right)
\end{aligned}$
&
$\left(\begin{array}{lll}0 & 0 & 0 \\ 0 & 0 & 0 \\ 0 & 0 & 0\end{array}\right) \left(\begin{array}{lll}0 & 0 & 0 \\ 0 & 0 & 0 \\ 0 & 0 & 0\end{array}\right)$ \\

\hline $\bar{3}{m}$ (${A}_{1g}$) & $\left(\begin{array}{lll}0 & 0 & 0 \\ 0 & 0 & 0 \\ 0 & 0 & 0\end{array}\right)$ & $\left(\begin{array}{lll}0 & 0 & 0 \\ 0 & 0 & 0 \\ 0 & 0 & 0\end{array}\right)$ & $\left(\begin{array}{ccc}0 & {A}_{12} & 0 \\ -{A}_{12} & 0 & 0 \\ 0 & 0 & 0\end{array}\right)$ & $\left(\begin{array}{lll}0 & 0 & 0 \\ 0 & 0 & 0 \\ 0 & 0 & 0\end{array}\right)$
&
$\begin{aligned} &
 \left(\begin{array}{ccc}0 & 0 & A_{113} \\ 0 & 0 & 0 \\ -A_{113} & 0 & 0\end{array}\right)
 \\&
\left(\begin{array}{ccc}0 & 0 & 0 \\ 0 & 0 & -A_{113} \\ 0 & A_{113} & 0\end{array}\right)
\end{aligned} $
& $\left(\begin{array}{lll}0 & 0 & 0 \\ 0 & 0 & 0 \\ 0 & 0 & 0\end{array}\right) \left(\begin{array}{lll}0 & 0 & 0 \\ 0 & 0 & 0 \\ 0 & 0 & 0\end{array}\right)$ \\
\hline
\end{tabular}
}
\end{table}
%%%%%%%%%%%%%%%%%%%%%%%%%%%%%%%%%%%%%%%%%%%%%

\begin{table}[H]
\centering
\caption{ The Raman tensors of the point group 6 ($C_6$) and corresponding MPGs.}
\label{TabC6}

\resizebox{\textwidth}{!}{%
%
}
\end{table}
%%%%%%%%%%%%%%%%%%%%%%%%%%%%%%%%%%%%%%%%%%%%%
\begin{table}[H]
\centering
\caption{ The Raman tensors of the point group $\bar{6}$ ($C_{3h}$) and corresponding MPGs.}
\label{TabC3h}

\resizebox{\textwidth}{!}{%
%
}
\end{table}
%%%%%%%%%%%%%%%%%%%%%%%%%%%%%%%%%%%%%%%%%%%%
\begin{table}[H]
\centering
\caption{ The Raman tensors of the point group $6/m$ ($C_{6h}$) and corresponding MPGs.}
\label{TabC6h}

\resizebox{\textwidth}{!}{%
\begin{tabular}{c|c|c|c|c|c|c|c|c|c|c|c|c}
\hline \hline
\rule{0pt}{12pt}
$6/m$ ($C_{6h}$) & ${A}_{{g}}$ & ${A}_{{u}}$ & ${B}_{{g}}$ & ${B}_{{u}}$ & ${ }^1 {E}_{1 {g}}$ & ${ }^2 {E}_{1 {g}}$ & ${ }^1 {E}_{1 {u}}$ & ${ }^2 {E}_{1 {u}}$ & ${ }^1 {E}_{2 {g}}$ & ${ }^2 {E}_{2 {g}}$ & ${ }^1 {E}_{2 {u}}$ & ${ }^2 {E}_{2 {u}}$ \\
\hline Sym. & $
\left(\begin{array}{ccc}
\alpha_{11} & 0 & 0 \\
0 & \alpha_{11} & 0 \\
0 & 0 & \alpha_{33}
\end{array}\right)
$ & $
\left(\begin{array}{lll}
0 & 0 & 0 \\
0 & 0 & 0 \\
0 & 0 & 0
\end{array}\right)
$ & $
\left(\begin{array}{lll}
0 & 0 & 0 \\
0 & 0 & 0 \\
0 & 0 & 0
\end{array}\right)
$ & $
\left(\begin{array}{lll}
0 & 0 & 0 \\
0 & 0 & 0 \\
0 & 0 & 0
\end{array}\right)
$ & $
\left(\begin{array}{ccc}
\alpha_{11} & i \alpha_{11} & 0 \\
i \alpha_{11} & -\alpha_{11} & 0 \\
0 & 0 & 0
\end{array}\right)
$ & $
\left(\begin{array}{ccc}
\alpha_{11} & -i \alpha_{11} & 0 \\
-i \alpha_{11} & -\alpha_{11} & 0 \\
0 & 0 & 0
\end{array}\right)
$ & $
\left(\begin{array}{lll}
0 & 0 & 0 \\
0 & 0 & 0 \\
0 & 0 & 0
\end{array}\right)
$ & $
\left(\begin{array}{lll}
0 & 0 & 0 \\
0 & 0 & 0 \\
0 & 0 & 0
\end{array}\right)
$ & $
\left(\begin{array}{ccc}
0 & 0 & \alpha_{13} \\
0 & 0 & -i \alpha_{13} \\
\alpha_{13} & -i \alpha_{13} & 0
\end{array}\right)
$
&
$\left(\begin{array}{ccc}0 & 0 & \alpha_{13} \\ 0 & 0 & i \alpha_{13} \\ \alpha_{13} & i \alpha_{13} & 0\end{array}\right)$
&
$ \left(\begin{array}{lll}
0 & 0 & 0 \\
0 & 0 & 0 \\
0 & 0 & 0
\end{array}\right)
$ & $
\left(\begin{array}{lll}
0 & 0 & 0 \\
0 & 0 & 0 \\
0 & 0 & 0
\end{array}\right)
$ \\
\hline \hline
\end{tabular}
}
\end{table}
%%%%%%%%%%%%%%%%%%%%%%%%%%%%%%%%%%%%%%%%%%%%%
\begin{table}[H]
\centering
\caption{ The Raman tensors of the point group 622 ($D_6$) and corresponding MPGs.}
\label{TabD6}

\resizebox{\textwidth}{!}{%
\begin{tabular}{c|c|c|c|c|c|c}
\hline \hline
\rule{0pt}{12pt}
 622 ($D_6$) & ${A}_1$ & ${A}_2$ & ${B}_{{1}}$ & ${B}_{{2}}$ & ${E}_1$ & ${E}_2$ \\
\hline Sym. & $\left(\begin{array}{ccc}\alpha_{11} & 0 & 0 \\ 0 & \alpha_{11} & 0 \\ 0 & 0 & \alpha_{33}\end{array}\right)$ & $\left(\begin{array}{lll}0 & 0 & 0 \\ 0 & 0 & 0 \\ 0 & 0 & 0\end{array}\right)$ & $\left(\begin{array}{lll}0 & 0 & 0 \\ 0 & 0 & 0 \\ 0 & 0 & 0\end{array}\right)$ & $\left(\begin{array}{lll}0 & 0 & 0 \\ 0 & 0 & 0 \\ 0 & 0 & 0\end{array}\right)$ &

$\begin{aligned} & \left(\begin{array}{ccc}0 & 0 & \alpha_{113} \\ 0 & 0 & 0 \\ \alpha_{113} & 0 & 0\end{array}\right) \\ & \left(\begin{array}{ccc}0 & 0 & 0 \\ 0 & 0 & \alpha_{113} \\ 0 & \alpha_{113} & 0\end{array}\right)\end{aligned}$
&
$
\begin{aligned}
& \left(\begin{array}{ccc}
\alpha_{111} & 0 & 0 \\
0 & -\alpha_{111} & 0 \\
0 & 0 & 0
\end{array}\right) \\
& \left(\begin{array}{ccc}
0 & \alpha_{111} & 0 \\
\alpha_{111} & 0 & 0 \\
0 & 0 & 0
\end{array}\right)
\end{aligned}
$ \\

\hline 622  ({${A}_1$})  & $\left(\begin{array}{lll}0 & 0 & 0 \\ 0 & 0 & 0 \\ 0 & 0 & 0\end{array}\right)$ & $\left(\begin{array}{ccc}0 & {A}_{12} & 0 \\ -{A}_{12} & 0 & 0 \\ 0 & 0 & 0\end{array}\right)$ & $\left(\begin{array}{lll}0 & 0 & 0 \\ 0 & 0 & 0 \\ 0 & 0 & 0\end{array}\right)$ & $\left(\begin{array}{lll}0 & 0 & 0 \\ 0 & 0 & 0 \\ 0 & 0 & 0\end{array}\right)$ &
$\begin{aligned} & \left(\begin{array}{ccc}0 & 0 & {A}_{113} \\ 0 & 0 & 0 \\ -{A}_{113} & 0 & 0\end{array}\right) \\ & \left(\begin{array}{ccc}0 & 0 & 0 \\ 0 & 0 & {A}_{113} \\ 0 & -{A}_{113} & 0\end{array}\right)\end{aligned}$
&
$\begin{aligned} & \left(\begin{array}{lll}0 & 0 & 0 \\ 0 & 0 & 0 \\ 0 & 0 & 0\end{array}\right) \\ & \left(\begin{array}{lll}0 & 0 & 0 \\ 0 & 0 & 0 \\ 0 & 0 & 0\end{array}\right)\end{aligned}$
\\
\hline \hline
\end{tabular}
}
\end{table}
%%%%%%%%%%%%%%%%%%%%%%%%%%%%%%%%%%%%%%%%%%%%%
\begin{table}[H]
\centering
\caption{ The Raman tensors of the point group $6mm$ ($C_{6v}$) and corresponding MPGs.}
\label{TabC6v}

\resizebox{\textwidth}{!}{%
\begin{tabular}{c|c|c|c|c|c|c}
\hline \hline
\rule{0pt}{12pt}
$6mm$ ($C_{6v}$) & ${A}_1$ & ${A}_2$ & ${B}_{{1}}$ & ${B}_{{2}}$ & ${E}_1$ & ${E}_2$ \\
\hline Sym. & $\left(\begin{array}{ccc}\alpha_{11} & 0 & 0 \\ 0 & \alpha_{11} & 0 \\ 0 & 0 & \alpha_{33}\end{array}\right)$ & $\left(\begin{array}{lll}0 & 0 & 0 \\ 0 & 0 & 0 \\ 0 & 0 & 0\end{array}\right)$ & $\left(\begin{array}{lll}0 & 0 & 0 \\ 0 & 0 & 0 \\ 0 & 0 & 0\end{array}\right)$ & $\left(\begin{array}{lll}0 & 0 & 0 \\ 0 & 0 & 0 \\ 0 & 0 & 0\end{array}\right)$ & $
\begin{gathered}
\left(\begin{array}{ccc}
0 & 0 & 0 \\
0 & 0 & \alpha_{123} \\
0 & \alpha_{123} & 0
\end{array}\right) \\
\left(\begin{array}{ccc}
0 & 0 & -\alpha_{123} \\
0 & 0 & 0 \\
-\alpha_{123} & 0 & 0
\end{array}\right)
\end{gathered}
$ & $
\begin{aligned}
& \left(\begin{array}{ccc}
\alpha_{111} & 0 & 0 \\
0 & -\alpha_{111} & 0 \\
0 & 0 & 0
\end{array}\right) \\
& \left(\begin{array}{ccc}
0 & \alpha_{111} & 0 \\
\alpha_{111} & 0 & 0 \\
0 & 0 & 0
\end{array}\right)
\end{aligned}
$ \\

\hline $6mm$ (${A}_1$) & $\left(\begin{array}{lll}0 & 0 & 0 \\ 0 & 0 & 0 \\ 0 & 0 & 0\end{array}\right)$ & $\left(\begin{array}{ccc}0 & {A}_{12} & 0 \\ -{A}_{12} & 0 & 0 \\ 0 & 0 & 0\end{array}\right)$ & $\left(\begin{array}{lll}0 & 0 & 0 \\ 0 & 0 & 0 \\ 0 & 0 & 0\end{array}\right)$ & $\left(\begin{array}{lll}0 & 0 & 0 \\ 0 & 0 & 0 \\ 0 & 0 & 0\end{array}\right)$ & $
\begin{aligned}
& \left(\begin{array}{ccc}
0 & 0 & 0 \\
0 & 0 & {A}_{123} \\
0 & -{A}_{123} & 0
\end{array}\right) \\
& \left(\begin{array}{ccc}
0 & 0 & -{A}_{123} \\
0 & 0 & 0 \\
{A}_{123} & 0 & 0
\end{array}\right)
\end{aligned}
$ & $
\begin{aligned}
& \left(\begin{array}{lll}
0 & 0 & 0 \\
0 & 0 & 0 \\
0 & 0 & 0
\end{array}\right) \\
& \left(\begin{array}{lll}
0 & 0 & 0 \\
0 & 0 & 0 \\
0 & 0 & 0
\end{array}\right)
\end{aligned}
$ \\
\hline \hline
\end{tabular}
}
\end{table}
%%%%%%%%%%%%%%%%%%%%%%%%%%%%%%%%%%%%%%%%%%%%%
%\clearpage  % »ò \cleardoublepage

\begin{table}[H]
\centering
\caption{ The Raman tensors of the point group $\bar{6}m2$ ($D_{3h}$) and corresponding MPGs.}
\label{TabD3h}
\resizebox{\textwidth}{!}{%
\begin{tabular}{c|c|c|c|c|c|c}
\hline \hline
\rule{0pt}{12pt}
$\bar{6}m2$ ($D_{3h}$) & $A_1^{\prime}$ & $A_1^{\prime \prime}$ & ${A}_2^{\prime}$ & ${A}_{{2}}^{{\prime} {\prime}}$ & ${E}^{\prime}$ & ${E}^{\prime \prime}$ \\
\hline Sym. & $\left(\begin{array}{ccc}\alpha_{11} & 0 & 0 \\ 0 & \alpha_{11} & 0 \\ 0 & 0 & \alpha_{33}\end{array}\right)$ & $\left(\begin{array}{lll}0 & 0 & 0 \\ 0 & 0 & 0 \\ 0 & 0 & 0\end{array}\right)$ &
$\left(\begin{array}{lll}0 & 0 & 0 \\ 0 & 0 & 0 \\ 0 & 0 & 0\end{array}\right)$ & $\left(\begin{array}{lll}0 & 0 & 0 \\ 0 & 0 & 0 \\ 0 & 0 & 0\end{array}\right)$ &
$\begin{aligned}
\left(\begin{array}{ccc}\alpha_{111} & 0 & 0 \\ 0 & -\alpha_{111} & 0 \\ 0 & 0 & 0\end{array}\right) \\
\left(\begin{array}{ccc}0 & -\alpha_{111} & 0 \\ -\alpha_{111} & 0 & 0 \\ 0 & 0 & 0\end{array}\right)
 \end{aligned}$
 &
$\begin{gathered}\left(\begin{array}{ccc}0 & 0 & \alpha_{113} \\ 0 & 0 & 0 \\ \alpha_{113} & 0 & 0\end{array}\right) \\ \left(\begin{array}{ccc}0 & 0 & 0 \\ 0 & 0 & -\alpha_{113} \\ 0 & -\alpha_{113} & 0\end{array}\right)\end{gathered}$
\\
\hline
$\bar{6}m2$  ($A_1^{\prime}$) & $\left(\begin{array}{lll}0 & 0 & 0 \\ 0 & 0 & 0 \\ 0 & 0 & 0\end{array}\right)$ & $\left(\begin{array}{lll}0 & 0 & 0 \\ 0 & 0 & 0 \\ 0 & 0 & 0\end{array}\right)$ & $\left(\begin{array}{ccc}0 & {A}_{12} & 0 \\ -{A}_{12} & 0 & 0 \\ 0 & 0 & 0\end{array}\right)$ & $\left(\begin{array}{lll}0 & 0 & 0 \\ 0 & 0 & 0 \\ 0 & 0 & 0\end{array}\right)$ & $\left(\begin{array}{lll}0 & 0 & 0 \\ 0 & 0 & 0 \\ 0 & 0 & 0\end{array}\right)$
$\left(\begin{array}{lll}0 & 0 & 0 \\ 0 & 0 & 0 \\ 0 & 0 & 0\end{array}\right)$
&
$\begin{aligned} & \left(\begin{array}{ccc}0 & 0 & {~A}_{113} \\ 0 & 0 & 0 \\ -{A}_{113} & 0 & 0\end{array}\right) \\ & \left(\begin{array}{ccc}0 & 0 & 0 \\ 0 & 0 & -{A}_{113} \\ 0 & {~A}_{113} & 0\end{array}\right)\end{aligned}$
 \\
\hline \hline
\end{tabular}
}
\end{table}
%%%%%%%%%%%%%%%%%%%%%%%%%%%%%%%%%%%%%%%%%%%%%
\begin{table}[H]
\centering
\caption{ The Raman tensors of the point group $6/{mmm}$ ($D_{6h})$  corresponding MPGs.}
\label{TabD6h}

\resizebox{\textwidth}{!}{%
%
}
\end{table}
%%%%%%%%%%%%%%%%%%%%%%%%%%%%%%%%%%%%%%%%%%%%%
\begin{table}[H]
\centering
\caption{ The Raman tensors of the point group 23 $(T)$ corresponding MPGs.}
\label{TabT}

\resizebox{\textwidth}{!}{%
%
}
\end{table}
%%%%%%%%%%%%%%%%%%%%%%%%%%%%%%%%%%%%%%%%%%%%%
\begin{table}[H]
\centering
\caption{ The Raman tensors of the point group $m\bar{3}$ ($T_h$) and corresponding MPGs.}
\label{TabTh}

\resizebox{\textwidth}{!}{%
\begin{tabular}{c|c|c|c|c|c|c|c|c}
\hline \hline
\rule{0pt}{12pt}
$m\bar{3}$ ($T_h$) & ${A}_{{g}}$ & ${A}_u$ & ${ }^1 {E}_{{g}}$ & ${ }^2 {E}_{{g}}$ & ${ }^1 {E}_u$ & ${ }^2 {E}_u$ & ${T}_{{g}}$ & ${T}_u$ \\
\hline Sym. & $\left(\begin{array}{ccc}\alpha_{11} & 0 & 0 \\ 0 & \alpha_{11} & 0 \\ 0 & 0 & \alpha_{11}\end{array}\right)$ & $\left(\begin{array}{lll}0 & 0 & 0 \\ 0 & 0 & 0 \\ 0 & 0 & 0\end{array}\right)$ & $\left(\begin{array}{ccc}\alpha_{11} & 0 & 0 \\ 0 & (-1)^{2 / 3} \alpha_{11} & 0 \\ 0 & 0 & -(-1)^{1 / 3} \alpha_{11}\end{array}\right)$ & $\left(\begin{array}{ccc}\alpha_{11} & 0 & 0 \\ 0 & -(-1)^{1 / 3} \alpha_{11} & 0 \\ 0 & 0 & (-1)^{2 / 3} \alpha_{11}\end{array}\right)$ & $\left(\begin{array}{lll}0 & 0 & 0 \\ 0 & 0 & 0 \\ 0 & 0 & 0\end{array}\right)$ & $\left(\begin{array}{lll}0 & 0 & 0 \\ 0 & 0 & 0 \\ 0 & 0 & 0\end{array}\right)$ & $
\begin{aligned}
& \left(\begin{array}{ccc}
0 & 0 & 0 \\
0 & 0 & \alpha_{123} \\
0 & \alpha_{123} & 0
\end{array}\right) \\
& \left(\begin{array}{ccc}
0 & 0 & \alpha_{123} \\
0 & 0 & 0 \\
\alpha_{123} & 0 & 0
\end{array}\right) \\
& \left(\begin{array}{ccc}
0 & \alpha_{123} & 0 \\
\alpha_{123} & 0 & 0 \\
0 & 0 & 0
\end{array}\right)
\end{aligned}
$ & $
\begin{aligned}
& \left(\begin{array}{lll}
0 & 0 & 0 \\
0 & 0 & 0 \\
0 & 0 & 0
\end{array}\right) \\
& \left(\begin{array}{lll}
0 & 0 & 0 \\
0 & 0 & 0 \\
0 & 0 & 0
\end{array}\right) \\
& \left(\begin{array}{lll}
0 & 0 & 0 \\
0 & 0 & 0 \\
0 & 0 & 0
\end{array}\right)
\end{aligned}
$ \\
\hline \hline
\end{tabular}
}
\end{table}
%%%%%%%%%%%%%%%%%%%%%%%%%%%%%%%%%%%%%%%%%%%%%
\begin{table}[H]
\centering
\caption{ The Raman tensors of the point group 432 ($O$) and corresponding MPGs.}
\label{TabO}

\resizebox{\textwidth}{!}{%
\begin{tabular}{c|c|c|c|c|c}
\hline \hline
\rule{0pt}{12pt}
432 ($O$) & ${A}_{{1}}$ & ${A}_2$ & $E$ & ${T}_1$ & ${T}_2$ \\
\hline Sym. & $\left(\begin{array}{ccc}\alpha_{11} & 0 & 0 \\ 0 & \alpha_{11} & 0 \\ 0 & 0 & \alpha_{11}\end{array}\right)$
&
$\left(\begin{array}{lll}0 & 0 & 0 \\ 0 & 0 & 0 \\ 0 & 0 & 0\end{array}\right)$
&
$\begin{gathered}\left(\begin{array}{cccc}\alpha_{111} & 0 & 0 \\ 0 & \alpha_{111} & 0 \\ 0 & 0 & -2  \alpha_{111}\end{array}\right) \\ \left(\begin{array}{ccc}-\sqrt{3} \alpha_{111} & 0 & 0 \\ 0 & \sqrt{3} \alpha_{111} & 0 \\ 0 & 0 & 0\end{array}\right)\end{gathered}$
& $
\begin{aligned}
& \left(\begin{array}{lll}
0 & 0 & 0 \\
0 & 0 & 0 \\
0 & 0 & 0
\end{array}\right) \\
& \left(\begin{array}{lll}
0 & 0 & 0 \\
0 & 0 & 0 \\
0 & 0 & 0
\end{array}\right) \\
& \left(\begin{array}{lll}
0 & 0 & 0 \\
0 & 0 & 0 \\
0 & 0 & 0
\end{array}\right)
\end{aligned}
$ & $
\begin{aligned}
& \left(\begin{array}{ccc}
0 & 0 & 0 \\
0 & 0 & \alpha_{123} \\
0 & \alpha_{123} & 0
\end{array}\right) \\
& \left(\begin{array}{ccc}
0 & 0 & \alpha_{123} \\
0 & 0 & 0 \\
\alpha_{123} & 0 & 0
\end{array}\right) \\
& \left(\begin{array}{ccc}
0 & \alpha_{123} & 0 \\
\alpha_{123} & 0 & 0 \\
0 & 0 & 0
\end{array}\right)
\end{aligned}
$ \\
\hline \hline
\end{tabular}
}
\end{table}
%%%%%%%%%%%%%%%%%%%%%%%%%%%%%%%%%%%%%%%%%%%%%
\begin{table}[H]
\centering
\caption{ The Raman tensors of the point group $\bar{4}3m$ ($T_d$) and corresponding MPGs.}
\label{TabTd}

\resizebox{\textwidth}{!}{%
\begin{tabular}{c|c|c|c|c|c}
\hline \hline
\rule{0pt}{12pt}
$\bar{4}3m$ ($T_d$) & ${A}_1$ & ${A}_2$ & E & ${T}_1$ & ${T}_2$ \\
\hline Sym. & $\left(\begin{array}{ccc}\alpha_{11} & 0 & 0 \\ 0 & \alpha_{11} & 0 \\ 0 & 0 & \alpha_{11}\end{array}\right)$ & $\left(\begin{array}{lll}0 & 0 & 0 \\ 0 & 0 & 0 \\ 0 & 0 & 0\end{array}\right)$
&

$\begin{gathered}\left(\begin{array}{cccc}\alpha_{111} & 0 & 0 \\ 0 & \alpha_{111} & 0 \\ 0 & 0 & -2 \alpha_{111}\end{array}\right) \\ \left(\begin{array}{ccc}-\sqrt{3} \alpha_{111} & 0 & 0 \\ 0 & \sqrt{3} \alpha_{111} & 0 \\ 0 & 0 & 0\end{array}\right)\end{gathered}$
&
$\begin{aligned}
& \left(\begin{array}{lll}
0 & 0 & 0 \\
0 & 0 & 0 \\
0 & 0 & 0
\end{array}\right) \\
& \left(\begin{array}{lll}
0 & 0 & 0 \\
0 & 0 & 0 \\
0 & 0 & 0
\end{array}\right) \\
& \left(\begin{array}{lll}
0 & 0 & 0 \\
0 & 0 & 0 \\
0 & 0 & 0
\end{array}\right)
\end{aligned}
$ & $
\begin{aligned}
& \left(\begin{array}{ccc}
0 & 0 & 0 \\
0 & 0 & \alpha_{123} \\
0 & \alpha_{123} & 0
\end{array}\right) \\
& \left(\begin{array}{ccc}
0 & 0 & \alpha_{123} \\
0 & 0 & 0 \\
\alpha_{123} & 0 & 0
\end{array}\right) \\
& \left(\begin{array}{ccc}
0 & \alpha_{123} & 0 \\
\alpha_{123} & 0 & 0 \\
0 & 0 & 0
\end{array}\right)
\end{aligned}
$ \\
\hline \hline
\end{tabular}
}
\end{table}
%%%%%%%%%%%%%%%%%%%%%%%%%%%%%%%%%%%%%%%%%%%%%

\begin{table}[H]
\centering
\caption{ The Raman tensors of the point group $m\bar{3}m$ ($O_h$) and corresponding MPGs.}
\label{TabOh}
\resizebox{\textwidth}{!}{%
\begin{tabular}{c|c|c|c|c|c|c|c|c|c|c}
\hline \hline
\rule{0pt}{12pt}
$m\bar{3}m$ ($O_h$) & ${A}_{{1 g}}$ & ${A}_{1 u}$ & ${A}_{{2}}{ }_{{g}}$ & ${A}_{2 u}$ & ${E}_{{g}}$ & ${E}_u$ & ${T}_{{1g}}$ & ${T}_{1u}$ & ${T}_{{2g}}$ & ${T}_{2u}$ \\
\hline \hline
\end{tabular}
}
\end{table}
%%%%%%%%%%%%%%%%%%%%%%%%%%%%%%%%%%%%%%%%%%%%%
%%%%%%%%%%%%%%%%%%%%%%%%%%%%%%%%%%%%%%%%%%%%%

\section{Compared with Corepresentation Method and Phenomenological Model Method}
\subsection{Corepresentation Method}

{For a magnetic point group \( M = S + AS \), where \( S \) is its unitary subgroup and \( AS \) is the anti-unitary coset, the Raman tensor is constrained on \( S \) by
\begin{equation}
R\boldsymbol{\alpha}^{(k)} = \mathbf{D}(R)\,\boldsymbol{\alpha}^{(k)}\,\mathbf{D}(R)^T = \sum_{p=1}^d \Gamma_{pk}^{(r)}(R)\,\boldsymbol{\alpha}^{(p)}, \quad R \in S.
\end{equation}
For an anti-unitary operation \( R' \), the constraint reads
\begin{equation}
R' \boldsymbol{\alpha}^{(k)} = \mathbf{D}(R)\,[\boldsymbol{\alpha}^{(k)}]^*\,\mathbf{D}(R)^T = \boldsymbol{\alpha}^{(k)},
\end{equation}
i.e., the time-reversal operation gives \( T\boldsymbol{\alpha}^{(k)} = [\boldsymbol{\alpha}^{(k)}]^* \), where $^*$ is the complex conjugate operation. Under these symmetry constraints, the Raman tensor elements can be purely real, purely imaginary, or complex. For example, for a gray group (nonmagnetic materials with time-reversal symmetry), the corepresentation theory predicts a purely real Raman tensor. On the other hand, no constraint is imposed on the relation between \( \alpha_{ij} \) and \( \alpha_{ji} \).}

{We find that this corepresentation method suffers from three main drawbacks:
\begin{enumerate}
    \item For nonmagnetic systems, the exchange symmetry \( \alpha_{ij} = \alpha_{ji} \) is not captured by the corepresentation approach.
    \item Since the Raman coefficients \( \alpha_{ij}(\omega) \) are functions of the photon energy \( \omega \), the Kramers-Kronig relations require that \( \alpha_{ij}(\omega) \) possesses both real and imaginary parts, analogous to the dielectric function \( \epsilon_{ij}(\omega) \). Thus, \( \alpha_{ij}(\omega) \) should generally be complex, except in the static limit \( \alpha_{ij}(0) \) for semiconductors. The corepresentation method fails to this requirement.
    \item The use of corepresentations to label phonon modes deviates from the conventional notation in practice.
\end{enumerate}}

{In our symmetry analysis method, the time-reversal operator $T$ exchanges the Raman tensor indices (\( ij \leftrightarrow ji \)) based on Onsager's reciprocity. Within this framework, the above requirements are naturally satisfied.}

\begin{table}[h]
\centering
\renewcommand{\arraystretch}{1.5}
\caption{Comparison between the corepresentation method and our method.}
\label{tab:method_compare}
\begin{tabular}{l|c|c}
\hline
\hline
 & \textbf{Corepresentation method} & \textbf{This work (Onsager reciprocity)} \\
\hline

Phonon notation & Corepresentation of magnetic point group \( M \) & Irrep of nonmagnetic point group \( G \) \\
\hline
Constraint of \( R' \) & \( R' \boldsymbol{\alpha}^{(k)} = \mathbf{D}(R)[\boldsymbol{\alpha}^{(k)}]^*\mathbf{D}(R)^T = \alpha^{(k)} \) & \( R' \boldsymbol{\alpha}^{(k)} = \mathbf{D}(R)[\boldsymbol{\alpha}^{(k)}]^T \mathbf{D}(R)^T = \sum_p \Gamma_{pk}^{(r)}(R)\boldsymbol{\alpha}^{(p)} \) \\
\hline
Raman coefficient & Purely real, imaginary or complex & Complex \\
\hline
Kramers-Kronig relation & \( \times \) & \( \checkmark \) \\
\hline
\hline
\end{tabular}
\end{table}

{A detailed comparison between the two methods is presented in Table \ref{tab:method_compare}. The Raman tensors obtained by the two methods for CrI$_3$ are presented in Table \ref{tab:tensor_compare}. We find that the results differ significantly. In particular, the corepresentation method restricts the Raman tensor elements to be purely real or purely imaginary, with no corrections to the off-diagonal elements. In contrast, our method allows each element to be complex, and imposes either symmetric or antisymmetric relations on the off-diagonal components. The selection rules derived from our approach are consistent with numerical calculations and successfully explain the experimental observations \cite{RN4012,RN4011}, whereas the corepresentation theory fails to do so.}

\begin{table}[h]
\centering
\caption{Comparison of Raman tensor forms obtained by the two methods for bilayer CrI$_3$. The letters (A--G) in the corepresentation method denote purely real quantities, while the Raman components in the last column are complex.}
\label{tab:tensor_compare}
\begin{tabular}{l|c|c}
\hline
\hline
\textbf{Magnetic point group} & \textbf{Corepresentation method} & \textbf{This work} \\

\hline
\( 2/m' \) (AFM, 2 $\parallel$ $y$)  &
 \begin{tabular}{c}
 $\boldsymbol{\alpha}(A) = \begin{pmatrix} A & 0 & B \\ 0 & E & 0 \\ D & 0 & I \end{pmatrix}$, \\ $\boldsymbol{\alpha}(B) = \begin{pmatrix} 0 & C & 0 \\ F & 0 & G \\ 0 & H & 0 \end{pmatrix} $ \end{tabular} &
  \begin{tabular}{c}
 $\boldsymbol{\alpha}(A_g) = \begin{pmatrix} \alpha_{11} & 0 & \boldsymbol{\alpha}_{13} \\ 0 & \alpha_{22} & 0 \\ \boldsymbol{\alpha}_{31} & 0 & \alpha_{33} \end{pmatrix},$ \\ $ \boldsymbol{\alpha}(B_u) = \begin{pmatrix} 0 & A_{12} & 0 \\ -A_{12} & 0 & A_{23} \\ 0 & -A_{23} & 0 \end{pmatrix} $ \end{tabular} \\
\hline
\( 2'/m' \) (FM, 2 $\parallel$ $y$) &
\begin{tabular}{c} $ \boldsymbol{\alpha}(A_g) = \begin{pmatrix} A & iC & B \\ iF & E & iG \\ D & iH & I \end{pmatrix},$ \\ $\boldsymbol{\alpha}(B_u) = 0 $ \end{tabular} &
\begin{tabular}{c} $\boldsymbol{\alpha}(A_g) = \begin{pmatrix} \alpha_{11} & A_{12} & \alpha_{13} \\ -A_{12} & \boldsymbol{\alpha}_{22} & A_{23} \\ \alpha_{31} & -A_{23} & \boldsymbol{\alpha}_{33} \end{pmatrix}, $ \\ $\boldsymbol{\alpha}(B_u) = 0 $ \end{tabular}  \\
\hline
\hline
\end{tabular}
\end{table}

\subsection{Phenomenological Model}

{The phenomenological model method reported in previous CrI$_3$ experimental studies \cite{RN4011,RN4012} is based on Placzek theory. For each phonon mode \(\mathbf{ U}_i \), its Raman tensor \( \boldsymbol{\alpha}^i \) is composed of the conventional structural contribution \( \boldsymbol{\alpha}_S^i \) and the layered-magnetism-assisted magnetic contribution \( \boldsymbol{\alpha}_{AS}^i \); i.e., \( \boldsymbol{\alpha}^i = \boldsymbol{\alpha}_S^i + \lambda_i \boldsymbol{\alpha}_{AS}^i \). The magneto-Raman tensor is antisymmetric, and \( \boldsymbol{\alpha}_{AS}^i \propto \mathbf{U}_i \cdot \mathbf{M} \) according to Placzek theory. \( \mathbf{M} = (\pm 1, \pm 1) \) indicates the spin moments pointing upward or downward, and \( \mathbf{U}_i \) denotes the interlayer vibration. Specifically, \( \mathbf{U}_1 = \frac{1}{\sqrt{2}}(1, 1) \) for the \( A_g \) mode, and \( \mathbf{U}_2 = \frac{1}{\sqrt{2}}(1, -1) \) for the \( B_u \) mode. Using the above relations, we obtain the Raman tensors for each configuration, which are also listed in Table \ref{tab:phenomenological} together with the corresponding vectors \( \mathbf{M} \) and \( \mathbf{U}_i \). }

\begin{table}[h]
\centering
\caption{Raman tensors of bilayer CrI$_3$ obtained from the phenomenological model method.}
\label{tab:phenomenological}
\begin{tabular}{l|c|c}
\hline
\hline
 & \textbf{AFM state, \( \mathbf{M} = (1, -1) \)} & \textbf{FM state, \( \mathbf{M} = (1, 1) \)} \\
\hline
\( A_g \) &
\begin{tabular}{c} $ \mathbf{U}_1 = \frac{1}{\sqrt{2}}(1, 1), \; \boldsymbol{\alpha}_{AS}^i = 0 $; \\
 $\boldsymbol{\alpha}(A_g) = \begin{pmatrix} \alpha_{11} & 0 & \alpha_{13} \\ 0 & \alpha_{22} & 0 \\ \alpha_{31} & 0 & \alpha_{33} \end{pmatrix} $ \end{tabular} &
\begin{tabular}{c} $ \mathbf{U}_1 = \frac{1}{\sqrt{2}}(1, 1), \; \boldsymbol{\alpha}_{AS}^i \neq 0 $; \\
 $\boldsymbol{\alpha}(A_g) = \begin{pmatrix} \alpha_{11} & A_{12} & \alpha_{13} \\ -A_{12} & \alpha_{22} & 0 \\ \alpha_{31} & 0 & \alpha_{33} \end{pmatrix}$ \end{tabular} \\
\hline
\( B_u \) &
\begin{tabular}{c} $ \mathbf{U}_2 = \frac{1}{\sqrt{2}}(1, -1), \; \boldsymbol{\alpha}_{AS}^i \neq 0 $;\\
$\boldsymbol{\alpha}(B_u) = \begin{pmatrix} 0 & A_{12} & 0 \\ -A_{12} & 0 & 0 \\ 0 & 0 & 0  \end{pmatrix}$ \end{tabular} &
\begin{tabular}{c} $ \mathbf{U}_2 = \frac{1}{\sqrt{2}}(1, -1), \; \boldsymbol{\alpha}_{AS}^i = 0 $; \\
$ \boldsymbol{\alpha}(B_u) = 0 $ \end{tabular}  \\
\hline
\hline
\end{tabular}
\end{table}

{Our approach successfully reproduces the phenomenological results. Besieds, our method further identifies an additional antisymmetric tensor component, \( \alpha_{23}^A = -\alpha_{32}^A \), for the \( B_u \) mode in the AFM state and for the \( A_g \) mode in the FM state, as shown in Fig. \ref{CrI3}. This component is small in magnitude and cannot be captured by previous phenomenological theories, as it stems from the $C_3$ symmetry breaking induced by interlayer stacking.}

\section{The characteristics of polarized Raman scattering}
\label{sec:AngleRaman}

% In Raman scattering, the contribution of a given Raman mode to the total signal intensity is given by $I^{(r)} \propto\left|\mathbf{e}_i \cdot \boldsymbol{\alpha}^{(r)} \cdot \mathbf{e}_s\right|^2$, where $\mathbf{e}_i$ and $\mathbf{e}_s$ are the polarization vectors of the incident and scattered light, respectively, and $\boldsymbol{\alpha}^{(r)}$ denotes the 3$\times$3 Raman tensor of mode $r$.

The contribution of a specific Raman mode to the Raman intensity is determined by
\begin{equation}
I\propto {{\left| \mathbf{e}_{s}^{\dagger }\cdot \boldsymbol{\alpha }\cdot \mathbf{e}_{i}^{T} \right|}^{2}},
\label{Eq-Raman_1}
\end{equation}
where $\mathbf{e}_i$ and $\mathbf{e}_s$ are the polarization vectors of the incident and scattered light, respectively. $\boldsymbol{\alpha}=\left[\alpha_{i j}\right]$ denotes the 3$\times$3 Raman tensor.
The Raman tensor in time-reversal symmetry-breaking systems can be decomposed into symmetric part
 $\left[\alpha_{i j}^S\right]$ and antisymmetric part $\left[\alpha_{i j}^A\right]$:
\begin{equation}
	\left[ {{\alpha }_{ij}} \right]=
\left[ \begin{array}{*{35}{l}}
   \alpha _{11} & \alpha _{12}& \alpha _{13} \\
   \alpha _{21} & \alpha _{22} & \alpha _{23}  \\
   \alpha _{31} & \alpha _{32} & \alpha _{33}  \\
\end{array} \right]=
\left[ \begin{array}{*{35}{l}}
   \alpha _{11}^{S} & \alpha _{12}^{S} & \alpha _{13}^{S}  \\
   \alpha _{12}^{S} & \alpha _{22}^{S} & \alpha _{23}^{S}  \\
   \alpha _{13}^{S} & \alpha _{23}^{S} & \alpha _{33}^{S}  \\
\end{array} \right]+
\left[ \begin{matrix}
   0 & \alpha _{12}^{A} & -\alpha _{31}^{A}  \\
   -\alpha _{12}^{A} & 0 & \alpha _{23}^{A}  \\
   \alpha _{31}^{A} & -\alpha _{23}^{A} & 0  \\
\end{matrix} \right].
\label{Eq-RamanTensor}
\end{equation}
The symmetric part of the Raman tensor ($\left[\alpha_{i j}^S\right]=\left[\alpha_{ji}^S\right]$) is even under time reversal, and invariant under the reversal of magnetic moment. In contrast, the antisymmetric part is odd under time reversal ($\left[\alpha_{i j}^A\right]=-\left[\alpha_{ji}^A\right]$), changing sign upon reversing the magnetic moment.

Depending on the forms of the Raman tensor, the phonon can be classified into four distinct scenarios: (1) purely nonmagnetic Raman activity, (2) purely magneto-Raman activity, (3) coexistence of both nonmagnetic and magneto-Raman activity, and (4) Raman inactivity. In the following, we will discuss the Raman features under the linearly-polarized and circularly-polarized light.
%The corresponding Raman spectrum characteristics are further discussed in Section IV of the Supplemental Material. The doubly degenerate $E$ mode will split into chiral phonons \cite{RN4026, RN4095, RN3942} in some MPGs, which are also discussed in detail in Section V of the Supplemental Material \cite{RN3775}.

\subsection{Linearly-polarized Raman scattering}

Under the parallel linear polarized light ($XX$), ${{\mathbf{e}}_{i}}={{\mathbf{e}}_{s}}=E(\cos \theta ,\sin \theta ,0)$, where $\theta$ is the angle between the polarized direction and the $x$-axis. According to Eq. (\ref{Eq-Raman_1}), the Raman intensity is given by:
\begin{equation}
I(XX)={{\alpha }_{11}}{{\cos }^{2}}\theta +{{\alpha }_{22}}{{\sin }^{2}}\theta +({{\alpha }_{12}}+{{\alpha }_{21}})\cos \theta \sin \theta .
\label{Eq-Raman2}
\end{equation}
Similarly, the Raman intensity under the cross-polarized light ($XY$) where ${{\mathbf{e}}_{i}}=E(\cos \theta ,\sin \theta ,0)$, ${{\mathbf{e}}_{s}}=E(-\sin \theta ,\cos \theta ,0)$ is
\begin{equation}
	I(XY)=(-{{\alpha }_{11}}+{{\alpha }_{22}})\cos \theta \sin \theta +{{\alpha }_{12}}{{\cos }^{2}}\theta -{{\alpha }_{21}}{{\sin }^{2}}\theta.
\label{Eq-Raman3}
\end{equation}
%%%%%%%%%%%%%%%%%%%%%%
The 1D magneto-Raman tensor often has the following forms:
\begin{equation}
	\left[ {{\alpha }_{ij}} \right]=
\left[ \begin{array}{*{35}{l}}
   \alpha _{11} & \alpha _{12}& 0 \\
   -\alpha _{12} & \alpha _{11} & 0  \\
   0 & 0 & \alpha _{33}  \\
\end{array} \right]
\label{Eq-RamanTensorAg}
\end{equation}

If the off-diagonal parts only has the anti-symmetric parts, \emph{i.e.} ${{\alpha }_{12}}=-{{\alpha }_{21}}$, and the diagonal parts are equal ${{\alpha }_{11}}={{\alpha }_{22}}$, we will have an angle-dependent Raman signal based on Eq. (\ref{Eq-Raman2}) and Eq. (\ref{Eq-Raman3}):
\begin{equation}
I({XX})\propto {{\left| {{\alpha }_{11}} \right|}^{2}},I({XY})\propto {{\left| {{\alpha }_{12}} \right|}^{2}}.
\label{Eq-Raman4}
\end{equation}

If we have the pure symmetric Raman elements, \emph{i.e.}, the off-diagonal parts are zeros, and diagonal parts exist and ${{\alpha }_{11}}={{\alpha }_{22}}$, we will have an angle-dependent Raman signal with these features
\begin{equation}
I({XX})\propto {{\left| {{\alpha }_{11}} \right|}^{2}},I({XY})=0.
\label{Eq-Raman5}
\end{equation}
Similarly, if only the pure magneto-Raman elements exist, \emph{i.e.}, the off-diagonal parts are anti-symmetric ${{\alpha }_{12}}=-{{\alpha }_{21}}$, and diagonal parts are zero, we will have an angle-dependent Raman signal:
\begin{equation}
I({XX})=0,I({XY})\propto {{\left| {{\alpha }_{12}} \right|}^{2}}.
\label{Eq-Raman6}
\end{equation}

\subsection{Circularly-polarized Raman scattering}

When we use circularly polarized light to detect the Raman spectrum, the left and right circular polarization vectors correspond to
\begin{equation}
\left\{ \begin{aligned}
  & {\boldsymbol{\sigma }_{L}}=\frac{1}{\sqrt{2}}(1,i,0), \\
 & {\boldsymbol{\sigma }_{R}}=\frac{1}{\sqrt{2}}(1,-i,0), \\
\end{aligned} \right.
\label{EqCircularRaman1}
\end{equation}
respectively. Depending on the helicity of the scattered and incident light, the Raman polarization setup can be $LL$ (left-left), $RR$ (right-right), $LR$ (left-right), and $RL$ (right-left) configurations. Therefore, the corresponding Raman intensities are
\begin{equation}
\left\{
\begin{aligned}
  & I({LL})\propto {{\left| {{\alpha }_{11}}+{{\alpha }_{22}}+i\left( {{\alpha }_{12}}-{{\alpha }_{21}} \right) \right|}^{2}}, \\
 & I({RR})\propto {{\left| {{\alpha }_{11}}+{{\alpha }_{22}}-i\left( {{\alpha }_{12}}-{{\alpha }_{21}} \right) \right|}^{2}}, \\
 & I({LR})\propto {{\left| {{\alpha }_{11}}-{{\alpha }_{22}}-i\left( {{\alpha }_{12}}+{{\alpha }_{21}} \right) \right|}^{2}}, \\
 & I({RL})\propto {{\left| {{\alpha }_{11}}-{{\alpha }_{22}}+i\left( {{\alpha }_{12}}+{{\alpha }_{21}} \right) \right|}^{2}}. \\
\end{aligned}
\label{EqCircularRaman2}
\right.
\end{equation}
Eq. (\ref{EqCircularRaman2}) can be further written as:
\begin{equation}
	\left\{
 \begin{aligned}
  & I(LL)\propto {{\left| \alpha _{11}^{S}+\alpha _{22}^{S}+i2\alpha _{12}^{A} \right|}^{2}}, \\
 & I(RR)\propto {{\left| \alpha _{11}^{S}+\alpha _{22}^{S}-i2\alpha _{12}^{A} \right|}^{2}}, \\
 & I(LR)\propto {{\left| \alpha _{11}^{S}-\alpha _{22}^{S}-i2\alpha _{12}^{S} \right|}^{2}}, \\
 & I(RL)\propto {{\left| \alpha _{11}^{S}-\alpha _{22}^{S}+i2\alpha _{12}^{S} \right|}^{2}}. \\
\end{aligned} \right.
\label{EqCircularRaman3}
\end{equation}

It is evident that in the $LR$ and $RL$ circular polarization channels, there are no anti-symmetric terms participating. If the anti-symmetric parts are zero, the Raman spectra have the feature: $I(LL)=I(RR)$. Therefore, the $LL$ and $RR$ setup can be used to detect the magneto-Raman effect. Moreover, a pivotal aspect to consider is that the Raman tensor elements must be complex numbers to ensure that $I(LL) \ne I(RR)$. If they are real, it will unavoidably result in $I(LL)=I(RR)$. This fact indicates that the corepresentation theory \cite{RN4022, RN4026} fails to capture the symmetry-constrained Raman tensors, and cannot explain the existing Raman experiments \cite{RN4012, RN4011, RN4025}.

A summary of the Raman intensity under both linearly and circularly polarized light is presented in Table \ref{TableRaman}. It is apparent that both the anti-symmetric and symmetric components of the Raman tensor exert a significant influence on the Raman spectrum under the linear and circular polarized Raman setup. Consequently, we can utilize these characteristics as criteria to distinguish and judge the contributions of different Raman tensor components.

\begin{table}[!htpb]
\centering
\caption{Raman intensity under the linear and circular polarized light. We assume that ${{\alpha }_{11}}={{\alpha }_{22}}$ if diagonal parts exist.}
\label{TableRaman}
\begin{tabular}{l|l|l}
\hline \hline
Raman tensor components & Linearly-polarized Raman & Circularly-polarized Raman \\
\hline
 \begin{tabular}{c}  Only the symmetric  \\   parts exist     \end{tabular}   & $
\begin{aligned}
& I({XX}) \propto\left|\alpha_{11}\right|^2 \\
& I({XY})=0
\end{aligned}
$ & $
\begin{aligned}
& I(L L)=I(R R) \neq 0 \\
& I(R L)=I(L R)=0
\end{aligned}
$ \\
\hline
 \begin{tabular}{c}  Only the antisymmetric  \\  parts exist   \end{tabular}   & $
\begin{aligned}
& I({XX})=0 \\
& I({XY}) \propto\left|\alpha_{12}\right|^2
\end{aligned}
$ & $
\begin{aligned}
& I(L L)=I(R R) \neq 0 \\
& I(R L)=I(L R)=0
\end{aligned}
$ \\
\hline
 \begin{tabular}{c}  Both the symmetric and  \\  anti-symmetric parts exist  \end{tabular}   & $
\begin{aligned}
& I({XX}) \propto\left|\alpha_{11}\right|^2 \\
& I({XY}) \propto\left|\alpha_{12}\right|^2
\end{aligned}
$ & $
\begin{aligned}
& I(L L) \neq I(R R) \neq 0 \\
& I(R L)=I(L R) \neq 0
\end{aligned}
$ \\
\hline \hline
\end{tabular}
\end{table}

\section{Raman scattering for the chiral phonons in MPGs with $C_3$ symmetry}
\label{sec:Chiarl}

In materials with $C_3$ symmetry (such as original point group 3, $\bar{3}$, 6, $\bar{6}$, and $6/m$, and black-white MPGs  $32'$, $3m'$, $\bar{6}m'2'$, $\bar{3}m'$, $62'2'$, $6m'm'$, and $6/mm'm'$), the characters of the irreps ${}^{1}E$ and ${}^{2}E$ are complex conjugate to each other: $\chi {{(}^{1}}E)={{\chi }^{*}}{{(}^{2}}E)$. Their basis functions are $x+iy$ and $x-iy$, respectively. When the time-reversal symmetry is broken, these basis functions correspond to a left-handed and a right-handed chiral vibrational mode, respectively, \emph{i.e.}, the chiral phonons. Their Raman tensors are:
\begin{equation}
\boldsymbol{\alpha }{{(}^{1}}E)=\left( \begin{matrix}
   {{\alpha }_{11}} & i{{\alpha }_{11}} & {{\alpha }_{13}}  \\
   i{{\alpha }_{11}} & -{{\alpha }_{11}} & -i{{\alpha }_{13}}  \\
   {{\alpha }_{13}} & -i{{\alpha }_{13}} & 0  \\
\end{matrix} \right),
\boldsymbol{\alpha }{{(}^{2}}E)=\left( \begin{matrix}
   {{{{\alpha }'}}_{11}} & -i{{{{\alpha }'}}_{11}} & {{{{\alpha }'}}_{13}}  \\
   -i{{{{\alpha }'}}_{11}} & -{{{{\alpha }'}}_{11}} & i{{{{\alpha }'}}_{13}}  \\
   {{{{\alpha }'}}_{13}} & i{{{{\alpha }'}}_{13}} & 0  \\
\end{matrix} \right).
\label{EqChiral1}
\end{equation}
It should be noted that the breaking of the time-reversal symmetry (such as ferromagnetic Co$_3$Sn$_2$S$_2$ \cite{RN3942}) may also casue the $^1E$ and $^2E$ Raman tensors in Eq. (\ref{EqChiral1}) not be complex conjugates of each other $\boldsymbol{\alpha }{{(}^{1}}E)\ne {{\left[ \boldsymbol{\alpha }{{(}^{2}}E) \right]}^{*}}$. Moreover, the non-diagonal elements of Eq. (\ref{EqChiral1}) are symmetric, while the Raman tensors [Eq. (\ref{Eq-RamanTensor})] of the 1D phonon induced by magnetism are asymmetric or antisymmetric.

Based on Eq. (\ref{EqCircularRaman2}) and (\ref{EqChiral1}), the Raman intensities of the chiral modes of ${}^{1}E$ mode under the circular light are given by
\begin{equation}
\left\{
\begin{aligned}
  & I({LL,}{}^{1}E)\propto {{\left| {{\alpha }_{11}}+{{\alpha }_{22}}+i\left( {{\alpha }_{12}}-{{\alpha }_{21}} \right) \right|}^{2}}=0, \\
 & I({RR,}{}^{1}E)\propto {{\left| {{\alpha }_{11}}+{{\alpha }_{22}}-i\left( {{\alpha }_{12}}-{{\alpha }_{21}} \right) \right|}^{2}}=0, \\
 & I(LR,{}^{1}E)\propto {{\left| {{\alpha }_{11}}-{{\alpha }_{22}}-i\left( {{\alpha }_{12}}+{{\alpha }_{21}} \right) \right|}^{2}}={{\left| 4{{\alpha }_{11}} \right|}^{2}}, \\
 & I(RL,{}^{1}E)\propto {{\left| {{\alpha }_{11}}-{{\alpha }_{22}}+i\left( {{\alpha }_{12}}+{{\alpha }_{21}} \right) \right|}^{2}}=0. \\
\end{aligned}
\label{EqChiral2}
\right.
\end{equation}
Similarly, the Raman intensities of the chiral phonon ${}^{2}E$ under the circular light are:
\begin{equation}
\left\{
\begin{aligned}
  & I({LL,}{}^{2}E)\propto {{\left| {{{{\alpha }'}}_{11}}+{{{{\alpha }'}}_{22}}+i\left( {{{{\alpha }'}}_{12}}-{{{{\alpha }'}}_{21}} \right) \right|}^{2}}=0, \\
 & I({RR,}{}^{2}E)\propto {{\left| {{{{\alpha }'}}_{11}}+{{{{\alpha }'}}_{22}}-i\left( {{{{\alpha }'}}_{12}}-{{{{\alpha }'}}_{21}} \right) \right|}^{2}}=0, \\
 & I(LR,{}^{2}E)\propto {{\left| {{{{\alpha }'}}_{11}}-{{{{\alpha }'}}_{22}}-i\left( {{{{\alpha }'}}_{12}}+{{{{\alpha }'}}_{21}} \right) \right|}^{2}}=0, \\
 & I(RL,{}^{2}E)\propto {{\left| {{{{\alpha }'}}_{11}}-{{{{\alpha }'}}_{22}}+i\left( {{{{\alpha }'}}_{12}}+{{{{\alpha }'}}_{21}} \right) \right|}^{2}}={{\left| 4{{{{\alpha }'}}_{11}} \right|}^{2}}. \\
\end{aligned} \right.
\label{EqChiral3}
\end{equation}
 Because $\boldsymbol{\alpha }{{(}^{1}}E)\ne {{\left[ \boldsymbol{\alpha }{{(}^{2}}E) \right]}^{*}}$,  $I(LR,{}^{1}E)\ne I(RL,{}^{2}E)$, \emph{i.e.} $LR$ and $RL$ have different peak heights \cite{RN3942}, as illustrated in Fig. \ref{Fig:RamanSpectrum}(a).

Given that the angular momenta of left-handed and right-handed light/chiral phonons possess values of $+\hbar $ and $-\hbar $ respectively, angular momentum conservation is fulfilled during this Raman scattering process [Eq. (\ref{EqChiral2}) and Eq. (\ref{EqChiral3})]. In contrast, the 1D magneto-Raman phonon in Section \ref{sec:AngleRaman} has no angular momentum, so $I(RL)= I(LR)= 0$, but $I(RR)\ne I(LL)\ne 0$ [Fig. \ref{Fig:RamanSpectrum}(b)]. Therefore, we can conclude that although the chiral phonons and 1D magneto-Raman phonons under circularly polarized light have phenomenologically similar Raman behavior, they have distinct underlying physics.

\begin{figure}[!htbp]
\centering
\includegraphics[width=0.9\textwidth]{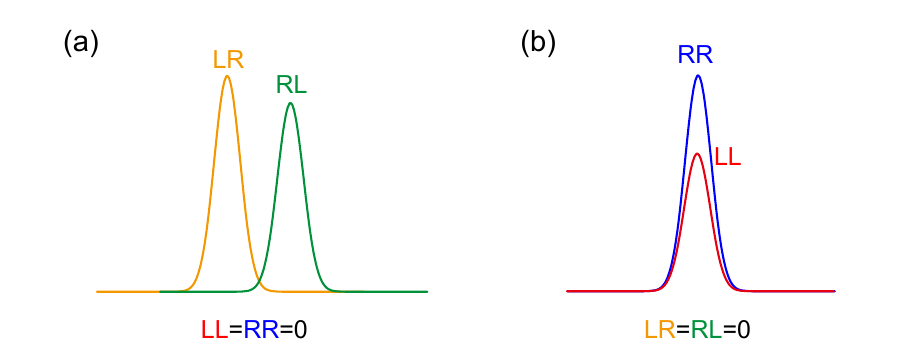}
\caption{Illustration of the Raman spectrum for (a) chiral phonons in magnetic materials and (b) one-dimensional magneto-Raman phonon. The Raman tensor in (a) is given by Eq. (\ref{EqChiral1}) with $\boldsymbol{\alpha }{{(}^{1}}E) \ne {{\left[ \boldsymbol{\alpha }{{(}^{2}}E) \right]}^{*}}$, and the Raman tensor in (b) follows Eq. (\ref{Eq-RamanTensorAg}).}
\label{Fig:RamanSpectrum}
\end{figure}

If the system exhibits time-reversal symmetry or other spatial symmetries, the ${}^{1}E$ and ${}^{2}E$ modes become degenerate, and $\boldsymbol{\alpha }{{(}^{1}}E)= {{\left[ \boldsymbol{\alpha }{{(}^{2}}E) \right]}^{*}}$. Consequently, they merge into two linearly-polarized modes along the $x$ and $y$ directions, respectively. Under such circumstances, the Raman tensors evolve into pure real matrices as shown below:
\begin{equation}
\left\{ \begin{matrix}
   \boldsymbol{\alpha }\left[E(x) \right]=\boldsymbol{\alpha }{{(}^{1}}E)+\boldsymbol{\alpha }{{(}^{2}}E)=2\left( \begin{matrix}
   {{\alpha }_{11}} & 0 & {{\alpha }_{13}}  \\
   0 & -{{\alpha }_{11}} & 0  \\
   {{\alpha }_{13}} & 0 & 0  \\
\end{matrix} \right)  \\
   \boldsymbol{\alpha }\left[ E(y) \right]=i\left[ \boldsymbol{\alpha }{{(}^{1}}E)-\boldsymbol{\alpha }{{(}^{2}}E) \right]=-2\left( \begin{matrix}
   0 & {{\alpha }_{11}} & 0  \\
   {{\alpha }_{11}} & 0 & {{\alpha }_{13}}  \\
   0 & {{\alpha }_{13}} & 0  \\
\end{matrix} \right)  \\
\end{matrix} \right.
\label{EqChiral4}
\end{equation}
Based on Eq. (\ref{EqCircularRaman2}), we find $I(LL)=I(RR)=0$, $I(LR)=I(RL)\ne 0$ for the two phonon modes $E(x)$ and $E(y)$.

It should be noted that in general textbooks and literature \cite{RN3992, RN3991}, Eq. (\ref{EqChiral4}) is frequently used, rather than Eq. (\ref{EqChiral1}). Nevertheless, the resulting Raman behavior fails to describe the experimental observations: only RL or LR Raman mode for every chiral phonon modes can be observed if the they are split. On the contrary, the Raman tensors in Eq. (\ref{EqChiral1}) and corresponding Raman behavior [Eq. (\ref{EqChiral2}) and Eq. (\ref{EqChiral3})] are consistent with the experimental observation \cite{RN3942, RN3992, RN3991}.

\section{First-Principles calculation method}

Under the independent-particle resonant Placzek approach illustrated in Fig. \ref{DFTworkflow} (a), the Raman coefficients $\alpha_{ij}^{(k)}$  of the $k$-th phonon mode can be obtained by taking the derivative of the dielectric susceptibility with respect to that normal mode coordinate $Q_k$ \cite{RN4092, RN4101, RN3941}:
\begin{equation}
\alpha _{ij}^{(k)}(\omega )={{\left. \frac{\partial {{\varepsilon }_{ij}}(\omega )}{\partial {{Q}_{k}}} \right|}_{Q=0}}\approx \frac{{{\varepsilon }_{ij}}(+\Delta ,\omega )-{{\varepsilon }_{ij}}(-\Delta ,\omega )}{2\Delta },
\label{Eq-Raman1}
\end{equation}
where $+\Delta$ and $-\Delta$ denote the maximum positive and negative displacement of the phonon mode, respectively.
Since the dielectric function tensor $\boldsymbol{\varepsilon}(\omega)$ is frequency dependent, the Raman tensor also relies on the incident light frequency $\omega $. Consequently, $\alpha_{ij}^{(k)}$ possesses both real and imaginary parts, analogous to the dielectric function, and satisfies the Kramers-Kronig relations. An important detail is that the spin-orbit coupling (SOC) effect should be considered in the dielectric function, as the antisymmetric part of ${{\varepsilon }_{ij}}(\omega )$ in collinear magnets vanishes without the SOC effect \cite{RN3983}. The first-principles calculation workflow to simulate the Raman effect is given in Fig. \ref{DFTworkflow}(b).
It should be noted that the independent-particle resonant Placzek approach suffers from the following limitations \cite{RN4131, RN4132, RN4133}: (1) absence of excitonic effects (electron-hole interaction); (2) absence of vertex corrections; (3) absence of full intermediate-state interference in the Kramers-Heisenberg-Dirac (KHD) formalism. Although this method cannot constitute a truly rigorous theory of resonant Raman scattering, it can capture certain resonant trends and symmetry.

\begin{figure}[!htbp]
\centering
\includegraphics[width=0.9\textwidth]{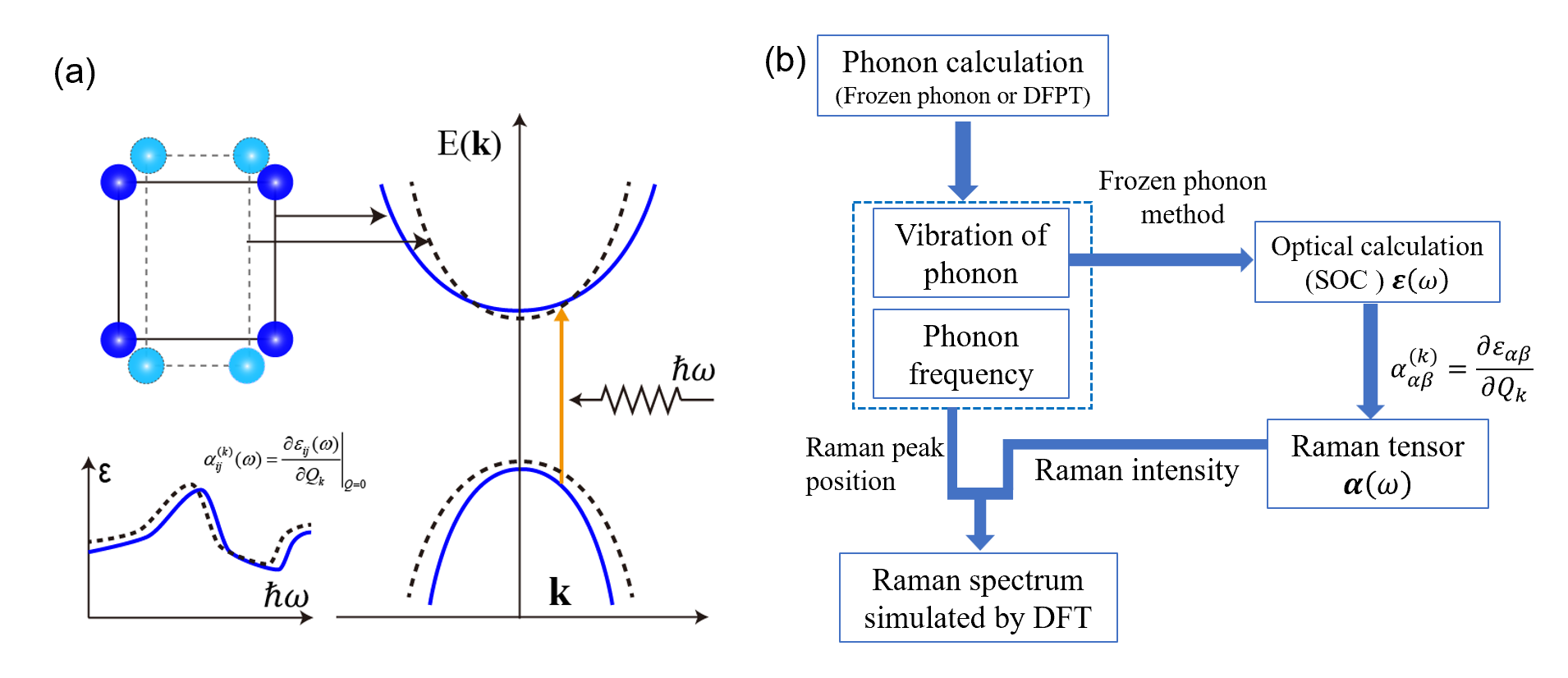}
\caption{(a) The illustration of the independent-particle resonant Placzek approach. (b) The calculation workflow of Raman properties based on first-principles calculation.}
\label{DFTworkflow}
\end{figure}

The first-principles calculations based on density functional theory (DFT) are performed using VASP software \cite{RN1434, RN1433}. The projector-augmented wave (PAW) method and generalized-gradient approximation (GGA) of the exchange-correlation energy functional in the Perdew-Burke-Ernzerhof (PBE) functional are employed; we further adapted the DFT-D2 functional to account for weak van der Waals (vdW) interlayer interactions for CrSBr and CrI$_3$. The spin-orbit coupling (SOC) effects are considered. $U_{eff}$= 3.0 eV was set for Cr/Mn atoms to account for strong electronic correlations. Phonon at the $\Gamma$ point ($\mathbf{q}=0$) is calculated with the density functional perturbation theory (DFPT) with the help of the Phonopy \cite{RN4097} package. The frequency-dependent dielectric function ${{\varepsilon }_{ij}}(\omega )$ is obtained based on the interband optical calculation (LOPTICS = .TRUE.).

\section{Calculated Raman tensor of bilayer CrI$_3$}

The monolayer CrI$_3$ has the $D_{3d}$ point group symmetry, while the bilayer CrI$_3$ has the $C_{2h}$ point group symmetry due to the interlayer stacking. The $A_g$ and $B_u$ modes of bilayer CrI$_3$ are in Fig. \ref{CrI3} (a, b), which mainly involve the vibration of the I atoms.
The calculated Raman coefficients in AFM and FM states are given in Fig. \ref{CrI3}(c-f).
In the AFM state, the $A_g$ mode only has the symmetric Raman component; and the $B_u$ mode, which is inactive in the paramagnetic state, has the magneto-Raman effect.
For the FM state, the $A_g$ mode has both the symmetric and antisymmetric parts, while the $B_u$ mode is inactive, as shown in Fig. \ref{CrI3} (e, f).
The above results are consistent with the symmetry analysis in the main text and experiments \cite{RN4012, RN4011, RN4025}.
Furthermore, we identify two additional antisymmetric tensor components ${{\alpha }_{23}}=-{{\alpha }_{32}}$ with very small magnitude in Fig. \ref{CrI3} (d,e), which cannot be captured by the generalized polarizability model. These tensor elements can be detected by the oblique incidence Raman setup in experiments.

\begin{figure}[!htbp]
\centering
\includegraphics[width=0.8\textwidth]{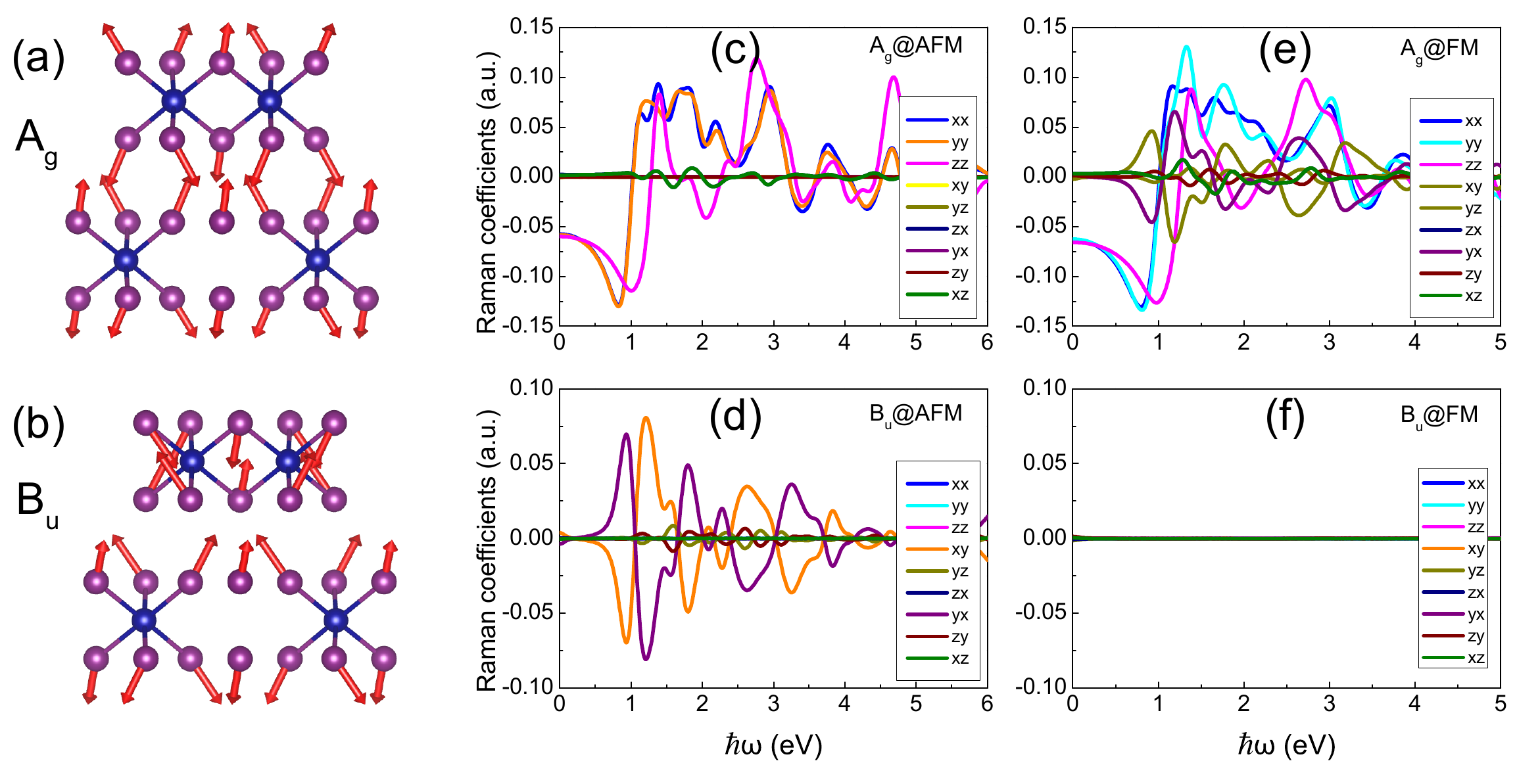}
\caption{ The vibration of (a) $A_g$ (125.4 cm$^{-1}$) and (b) $B_u$ (121.4 cm$^{-1}$) modes of bilayer CrI$_3$. (c, e)/(d, f) the real part of the Raman coefficients of $A_g$ and $B_u$ modes in AFM/FM states with light energy. The imaginary part can be obtained by the Kramers-Kronig relations.}
\label{CrI3}
\end{figure}

\section{Calculated Raman tensor of bilayer CrSBr}
CrSBr has a layered A-type magnetic structure. Its parent space group is $Pmnm$ (No. 59), and the corresponding point group is ${D}_{2h}({mmm})$. Each layer contains 6 atoms. Therefore, there are 18 vibrational modes:
\begin{equation}
\begin{gathered}
\Gamma_{\text{acoustic }}=B_{1u}+B_{2u}+B_{3u}, \\
\Gamma_{\text{optic }}=3 A_g+3 B_{2g}+3 B_{3g}+2 B_{1u}+2 B_{2u}+2 B_{3u}.
\end{gathered}
\end{equation}
Among them, ${A}_{{g}}$, ${B}_{2{g}}$, and ${B}_{3{g}}$ are Raman-active modes, whereas ${B}_{1{u}},{B}_{2{u}}$, and ${B}_{3{u}}$ are infrared-active modes. The active Raman tensors without considering the magnetic state are:
\begin{equation}
\boldsymbol{\alpha} \left(A_g\right)=\left(\begin{array}{lll}
a & 0 & 0 \\
0 & b & 0 \\
0 & 0 & c
\end{array}\right),
\boldsymbol{\alpha}\left(B_{2 g}\right)=\left(\begin{array}{lll}
0 & 0 & d \\
0 & 0 & 0 \\
d & 0 & 0
\end{array}\right),
\boldsymbol{\alpha} \left(B_{3 g}\right)=\left(\begin{array}{lll}
0 & 0 & 0 \\
0 & 0 & e \\
0 & e & 0
\end{array}\right).
\label{EqCrSBr}
\end{equation}
Since the in-plane tensor components of ${B}_{2{g}}$ and ${B}_{3{g}}$ are not present in Eq. (\ref{EqCrSBr}), these modes is deemed undetectable. However, in the monolayer and odd layer cases, the MPG is $m' mm'$. According to Table \ref{TabD2h}, the $B_{3g}$ phonon mode has the in-plane magneto-Raman elements due to the existing anti-symmetric Raman elements ${{\alpha }_{12}}=-{{\alpha }_{21}}$.

In a bilayer, the $B_{3g}$ phonon in monolayer splits into an even-parity $B_{3g}$ mode and an odd-parity $B_{2u}$ mode due to Davydov splitting, as shown in Fig. \ref{FigCrSBr} (a, b). These two modes have nearly degenerate frequencies (358.73 cm$^{-1}$ and 358.74 cm$^{-1}$ in our calculation), due to the weak interlayer coupling. In the bilayer CrSBr with AFM state, its MPG is $m' mm$ (corresponding to $B_{3u}$). The $B_{3g}$ phonon mode has no antisymmetric part, because ${{B}_{3g}}\otimes {{B}_{3u}}={{A}_{u}}$, and  $A_{u} $ is not the irrep of axial vectors. In contrast, the $B_{2u}$ phonon mode acquires magneto-Raman activity, because ${{B}_{2u}}\otimes {{B}_{3u}}={{B}_{1g}}$. $B_{1g}$ is the irrep of axial vector compounds ${{J}_{z}}$, which allows the $\boldsymbol{\alpha}^{A}$ with ${{\alpha }_{12}}=-{{\alpha }_{21}}$, according to Table \ref{TabD2h}. These symmetry analysis results are verified by our numerical calculations, as shown in Fig. \ref{FigCrSBr} (f).

In the bulk CrSBr, the magnetic space group is $P_anma$, which has the effective time-reversal symmetry. This symmetry leads to the absence of the magneto-Raman effect according to our theory, which is consistent with experiments \cite{RN4038, RN4042}.

\begin{figure}[!htbp]
\centering
\includegraphics[width=0.8\textwidth]{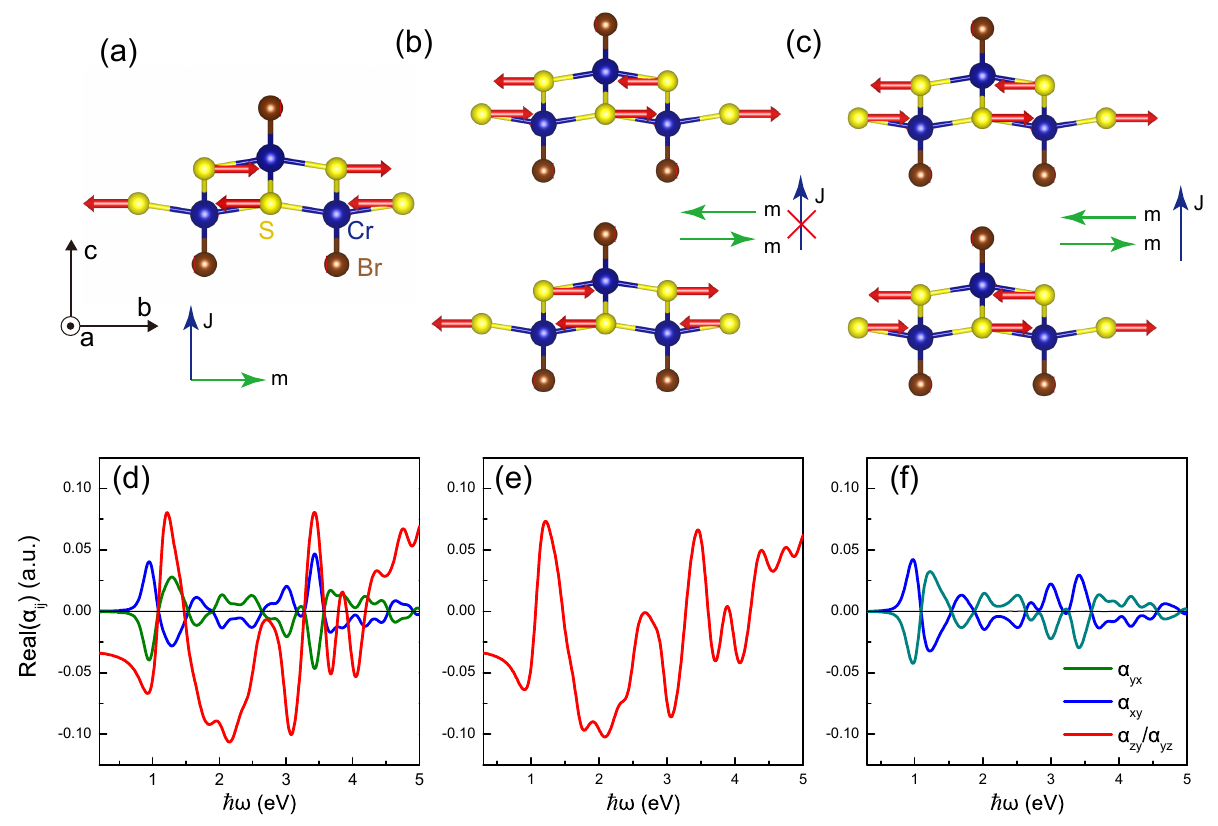}
\caption{   The vibrational modes of (a) $B_{3g}$ in monolayer and corresponding Davydov splitting modes (b) $B_{3g}$ and (c) $B_{2u}$ in bilayer.
 (d-f) The calculated real part of the Raman susceptibilities for the modes in (a-c). The imaginary part can be obtained by the Kramers-Kronig relations.}
\label{FigCrSBr}
\end{figure}

\section{Raman characteristics of NiAs-type altermagnetic materials}
\label{sec:NiAs}
Hexagonal NiAs-type altermagnetic materials include $\alpha$-MnTe, CrSb, FeS, \emph{etc.}. The parent space group of these magnetic materials is $P6_3/mmc$ (No. 194). In these materials, the two magnetic atoms with opposite spins occupy $2a$ sites at (0, 0, 0) and (0, 0, 1/2), and the nonmagnetic atoms are located at the $2d$ sites (1/3, 2/3, 3/4) and (2/3, 1/3, 1/4). Correspondingly, the phonon modes at the $\Gamma$ point are
\begin{equation}
{{\Gamma }_{{optic }}}={{A}_{2u}}\oplus {{B}_{2g}}\oplus {{B}_{2u}}\oplus {{E}_{2u}}\oplus {{E}_{2g}}\oplus {{E}_{1u}}.
\label{EqNiAs1}
\end{equation}
As shown in Fig. \ref{FigNiAsPhonon}, the $B_{2g}$ phonon mode corresponds to the symmetric vibration of non-magnetic atoms along the $z$-direction, while the $E_{2g}$ mode involves the vibration of these two atoms along the $x$ and $y$ directions, respectively.

\begin{figure}[!htbp]
\centering
\includegraphics[width=0.8\textwidth]{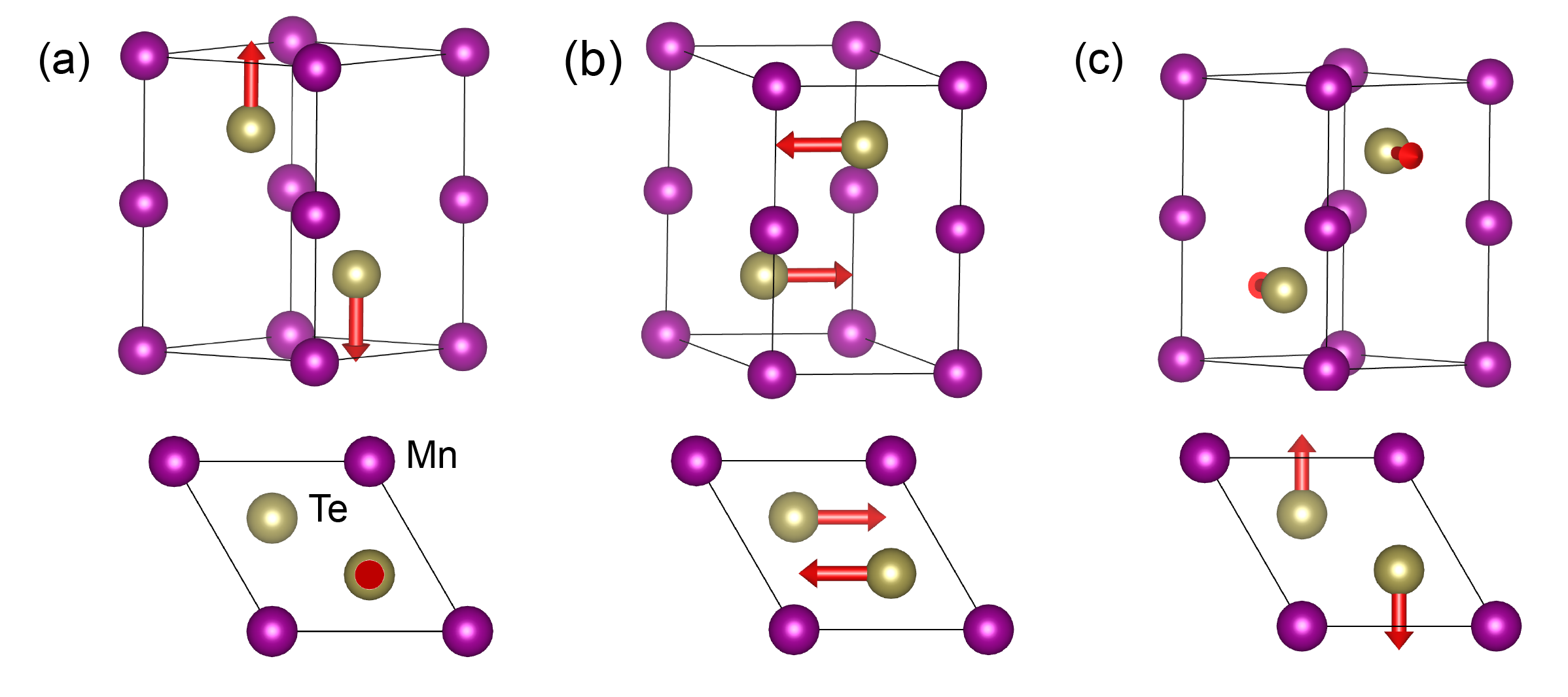}
\caption{ The phonon modes of (a) $B_{2g}$ and (b) $E_{2g} (x)$ and (c) $E_{2g} (y)$ of the NiAs-type altermagnets. These modes will have the following evolution: $B_{2g}\to B_{3g}$, $E_{2g} (x)\to B_{1g}$, $E_{2g} (y) \to A_g$ from the $D_{6h}$ to $D_{2h}$. The magnetic atoms at the (0,0,0) and (0, 0, 1/2) have the opposite spin.}
\label{FigNiAsPhonon}
\end{figure}

If the magnetic moment is aligned along the $z$-axis, the MPG is $6'/m' m' m$, and this MPG corresponds to $B_{1g}$ representation of $D_{6h}$ ($6/mmm$). According to Table \ref{TabD6h}, the $B_{2g}$ mode, which is Raman inactive in the nonmagnetic state, exhibits magneto-Raman activity. Besides, the $E_{2g}$ mode is intrinsically Raman active without magnetism and can possess additional magneto-Raman elements. Their Raman tensors are as follows:
\begin{equation}
\boldsymbol{\alpha} \left( {{B}_{2g}} \right)=0+\left( \begin{matrix}
   0 & {{A}_{12}} & 0  \\
   -{{A}_{12}} & 0 & 0  \\
   0 & 0 & 0  \\
\end{matrix} \right),
\label{EqNiAs2}
\end{equation}
\begin{equation}
\boldsymbol{\alpha} ({{E}_{2g}},x)=\left( \begin{matrix}
   0 & -{{\alpha }_{11}} & 0  \\
   -{{\alpha }_{11}} & 0 & 0  \\
   0 & 0 & 0  \\
\end{matrix} \right)+\left( \begin{matrix}
   0 & 0 & 0  \\
   0 & 0 & {{A}_{13}}  \\
   0 & -{{A}_{13}} & 0  \\
\end{matrix} \right),
\label{EqNiAs3}
\end{equation}
\begin{equation}
\boldsymbol{\alpha} ({{E}_{2g}},y)=\left( \begin{matrix}
   {{\alpha }_{11}} & 0 & 0  \\
   0 & -{{\alpha }_{11}} & 0  \\
   0 & 0 & 0  \\
\end{matrix} \right)+\left( \begin{matrix}
   0 & 0 & {{A}_{13}}  \\
   0 & 0 & 0  \\
   -{{A}_{13}} & 0 & 0  \\
\end{matrix} \right).
\label{EqNiAs4}
\end{equation}

If the Ne\'{e}l vector is aligned along the $x$-axis, the MPG is the original point group $mmm$, which corresponds to the $A_g$ irrep of $D_{2h}$. According to the compatibility relations, the $B_{2g}$ phonon in the $D_{6h}$ group transforms into the $B_{3g}$ of $D_{2h}$. Based on Table \ref{TabD2h}, its corresponding Raman tensor is:
\begin{equation}
\boldsymbol{ \alpha} ({{B}_{3g}})=\left( \begin{matrix}
   0 & 0 & 0  \\
   0 & 0 & {{\alpha }_{23}}  \\
   0 & {{\alpha }_{23}} & 0  \\
\end{matrix} \right)+\left( \begin{matrix}
   0 & 0 & 0  \\
   0 & 0 & {{A}_{23}}  \\
   0 & -{{A}_{23}} & 0  \\
\end{matrix} \right),
\label{EqNiAs5}
\end{equation}
\emph{i.e.}, this mode exhibits magneto-Raman activity in both even-parity [$\boldsymbol{\alpha}^S$] and odd-parity parts [$\boldsymbol{\alpha}^A$]. The $E_{2g}$ mode along the $x$ and $y$ directions in the $D_{6h}$ group splits into $B_{1g}$ and $A_g$ modes under $D_{2h}$. The corresponding Raman tensors are as follows:
\begin{equation}
\boldsymbol{ \alpha } ({{B}_{1g}})=\left( \begin{matrix}
   0 & {{\alpha }_{12}} & 0  \\
   {{\alpha }_{12}} & 0 & 0  \\
   0 & 0 & 0  \\
\end{matrix} \right)+\left( \begin{matrix}
   0 & {{A}_{12}} & 0  \\
   -{{A}_{12}} & 0 & 0  \\
   0 & 0 & 0  \\
\end{matrix} \right),
\label{EqNiAs6}
\end{equation}
\begin{equation}
\boldsymbol{ \alpha} ({{A}_{g}})=\left( \begin{matrix}
   {{\alpha }_{11}} & 0 & 0  \\
   0 & {{\alpha }_{22}} & 0  \\
   0 & 0 & {{\alpha }_{33}}  \\
\end{matrix} \right)+0.
\label{EqNiAs7}
\end{equation}
%According to Eqs. (\ref{EqNiAs5})-(\ref{EqNiAs7}), the Raman tensors of $B_{3g}$ and $B_{1g}$ modes contain both symmetric and antisymmetric parts, whereas the Raman tensor of $A_g$ mode only consists of the symmetric part.

\begin{table}[!tb]
\centering
\caption{The compatibility and the Raman tensor of the NiAs-type altermagnets with different Ne\'{e}l directions.}
\label{TabNiAs}
\resizebox{\textwidth}{!}{%
\begin{tabular}{c|c|c|c}
\hline \hline
\backslashbox{Ne\'{e}l vector \& MPG }{Phonon irrep}
& ${B}_{{2 g}}$ & ${E}_{{2 g}} {( x )}$ & ${E}_{{2 g}} { ( y )}$ \\
\hline $z$ ($6'/m'm'm$, $B_{1g}$) &
$0+\left(\begin{array}{ccc}
0 & A_{12} & 0 \\
-A_{12} & 0 & 0 \\
0 & 0 & 0 \end{array}\right)$
& $\left(\begin{array}{ccc}0 & -\alpha_{11} & 0 \\ -\alpha_{11} & 0 & 0 \\ 0 & 0 & 0\end{array}\right)+\left(\begin{array}{ccc}0 & 0 & 0 \\ 0 & 0 & A_{13} \\ 0 & -A_{13} & 0\end{array}\right)$ & $\left(\begin{array}{ccc}\alpha_{11} & 0 & 0 \\ 0 & -\alpha_{11} & 0 \\ 0 & 0 & 0\end{array}\right)+\left(\begin{array}{ccc}0 & 0 & A_{13} \\ 0 & 0 & 0 \\ -A_{13} & 0 & 0\end{array}\right)$ \\
\hline $x$ ($mmm$, $A_g$) & \begin{tabular}{c} ${B}_{3 {g}}$ \\
$\left(\begin{array}{ccc}
0 & 0 & 0 \\
0 & 0 & \alpha_{23} \\
0 & \alpha_{23} & 0
\end{array}\right)+\left(\begin{array}{ccc}
0 & 0 & 0 \\
0 & 0 & A_{23} \\
0 & -A_{23} & 0
\end{array}\right) $ \end{tabular}
 &
 \begin{tabular}{c}  ${B}_{1 {g}}$ \\
$\left(\begin{array}{ccc}
0 & \alpha_{12} & 0 \\
\alpha_{12} & 0 & 0 \\
0 & 0 & 0
\end{array}\right)+\left(\begin{array}{ccc}
0 & A_{12} & 0 \\
-A_{12} & 0 & 0 \\
0 & 0 & 0
\end{array}\right)$  \end{tabular}
& $\begin{gathered}
{A}_{{g}} \\
\left(\begin{array}{ccc}
\alpha_{11} & 0 & 0 \\
0 & \alpha_{22} & 0 \\
0 & 0 & \alpha_{33}
\end{array}\right)+0
\end{gathered}
$ \\
\hline
$y$ ($m'm'm$, $B_{1g}$) &
\begin{tabular}{c} ${B}_{3 {g}}$ \\
$\left(\begin{array}{ccc}
0 & 0 & 0 \\
0 & 0 & \alpha_{23} \\
0 & \alpha_{23} & 0
\end{array}\right)+\left(\begin{array}{ccc}
0 & 0 & -A_{31} \\
0 & 0 & 0 \\
A_{31} & 0 & 0
\end{array}\right)$ \end{tabular} &
\begin{tabular}{c} ${B}_{1 {g}}$ \\
$\left(\begin{array}{ccc}
0 & \alpha_{12} & 0 \\
\alpha_{12} & 0 & 0 \\
0 & 0 & 0
\end{array}\right)+0
$ \end{tabular}
&
\begin{tabular}{c} $A_g$ \\
$\left(\begin{array}{ccc}
\alpha_{11} & 0 & 0 \\
0 & \alpha_{22} & 0 \\
0 & 0 & \alpha_{33}
\end{array}\right)+\left(\begin{array}{ccc}
0 & A_{12} & 0 \\
-A_{12} & 0 & 0 \\
0 & 0 & 0
\end{array}\right)$ \end{tabular}
\\
\hline \hline
\end{tabular}
}
\end{table}

If the Ne\'{e}l vector is aligned along the $y$-axis, the MPG is $m' m' m$, corresponding to $B_{1g}$ irrep of  $D_{2h}$. Similarly, the $B_{2g}$ phonon in the $D_{6h}$ group transforms into the $B_{3g}$  irrep of $D_{2h}$. The corresponding Raman tensor is given by:
\begin{equation}
\boldsymbol{\alpha} ({{B}_{3g}})=\left( \begin{matrix}
   0 & 0 & 0  \\
   0 & 0 & {{\alpha }_{23}}  \\
   0 & {{\alpha }_{23}} & 0  \\
\end{matrix} \right)+\left( \begin{matrix}
   0 & 0 & -{{A}_{31}}  \\
   0 & 0 & 0  \\
   {{A}_{31}} & 0 & 0  \\
\end{matrix} \right).
\label{EqNiAs8}
\end{equation}
The Raman tensor for the $B_{1g}$ (from the irrep $E_{2g}(x)$ of $D_{6h}$) is:
\begin{equation}
\boldsymbol{\alpha} ({{B}_{1g}})=\left( \begin{matrix}
   0 & {{\alpha }_{12}} & 0  \\
   {{\alpha }_{12}} & 0 & 0  \\
   0 & 0 & 0  \\
\end{matrix} \right)+0.
\label{EqNiAs9}
\end{equation}
And the Raman tensor for the mode vibrating along the $y$-direction ($A_g$) (from the irrep $E_{2g}(y)$ of $D_{6h}$) is:
\begin{equation}
\boldsymbol{\alpha} ({{A}_{g}})=\left( \begin{matrix}
   {{\alpha }_{11}} & 0 & 0  \\
   0 & {{\alpha }_{22}} & 0  \\
   0 & 0 & {{\alpha }_{33}}  \\
\end{matrix} \right)+\left( \begin{matrix}
   0 & {{A}_{12}} & 0  \\
   -{{A}_{12}} & 0 & 0  \\
   0 & 0 & 0  \\
\end{matrix} \right).
\label{EqNiAs10}
\end{equation}
Finally, the Raman tensors for the magnetic moment along the above three Ne\'{e}l directions are summarized in Table \ref{TabNiAs}.

\begin{figure}[!htbp]
\centering
\includegraphics[width=0.85\textwidth]{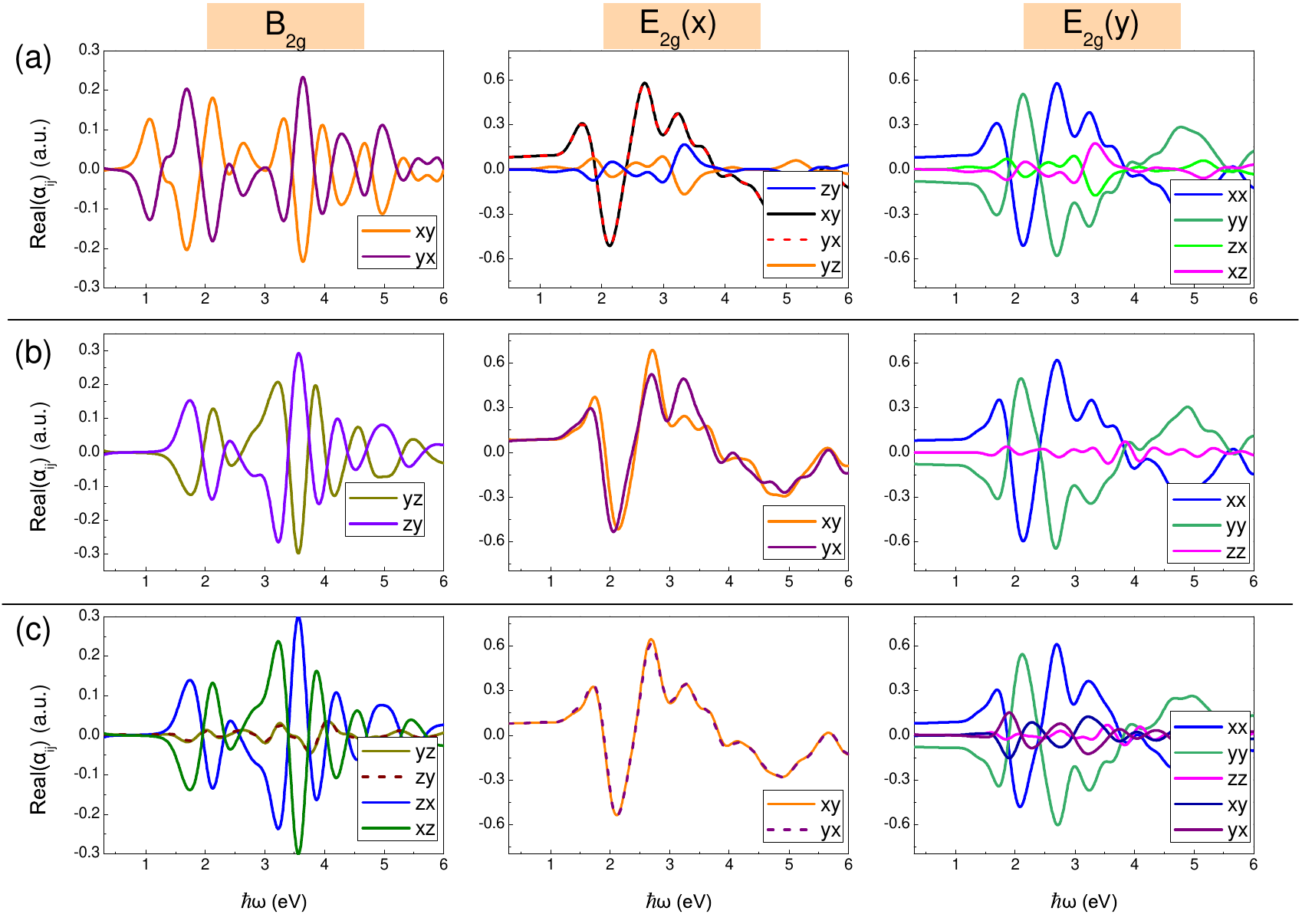}
\caption{Calculated real part of the Raman coefficients for NiAs-type altermagnet MnTe for the $B_{2g}$ and $E_{2g}$ modes when the magnetic moment is aligned along the (a) $z$, (b) $x$, and (c) $y$ axes, respectively. The imaginary part can be obtained by the Kramers-Kronig relations. The other Raman elements that are not displayed are zeros forced by symmetry.}
\label{FigNiAs}
\end{figure}

The first-principles calculation results of the Raman coefficients for the three magnetic moment directions are shown in Fig. \ref{FigNiAs}, which are consistent with the above symmetry analysis. Furthermore, we find that the effective vector of the Raman matrix induced by these vibrational modes exhibits a rather complex relationship with both the vibrational directions and the magnetic moment directions. %For instance, the Raman axial vector moment is parallel to the Ne\'{e}l vector for the arising from vibration modes along the $z$-axis (the second column in Table \ref{TabNiAs}). However, the Raman vector of the $E_g$ mode is perpendicular to the Ne\'{e}l vector.
This stems from the fact that the magneto-Raman tensor originates from the direct product of the irrep of the MPG ($\chi_{mag}$) and the phonon ($\Gamma^{(r)}$).

Our symmetry analysis indicates that the one-dimensional vibrational mode $B_{2g}$ at 122 cm$^{-1}$ of MnTe can exhibit magnetically induced Raman activity, providing a possible explanation for previous Raman experiments. We also note that { the persistence of this Raman signal up to 400 K \cite{RN4128, RN4126} in experiments, well above the N\'eel temperature ($T_N$=307 K), suggests that it is unlikely to be induced solely by long-range magnetic order. One possible explanation is that the observed signal is dominated by an extrinsic mechanism (such as surface \cite{RN4124,RN4125}, impurity, or plasmon states \cite{RN4129}), while the magneto-Raman contribution is too weak to be resolved even below $T_N$.} Besides, its Raman spectrum remains controversial \cite{RN4127, RN4129}. Therefore, more detailed theoretical and experimental measurements are still required.
{This example is included primarily as a demonstration of symmetry analysis-illustrating how our theoretical framework constrains the Raman tensor forms and their transformation relations when the magnetic moment is oriented along different directions-rather than as direct experimental validation of the magneto-Raman effect in this specific material.}

\section{Geometric relation similarity of magneto-Raman effect and anomalous Hall effect in altermagnets }

{It have been shown that in the altermagnet, the Hall vector (or magneto-optical vector) is always orthogonal to the N\'{e}el vector \cite{RN4079,RN4045,RN3940}. This phenomenon has been further analyzed from a symmetry perspective in Ref. \cite{RN3940}: the transformation properties of the N\'{e}el vector $\mathbf{n}$ and the Hall vector $\boldsymbol{\sigma}_H$ differ under symmetry operations. Moreover, Refs. \cite{RN3736,RN3918} employed Landau theory and group representation methods to give a more mathematical explanation. The Hall vector $\boldsymbol{\sigma}_H$ belongs to the axial representation $\Gamma_A$, while the N\'{e}el vector $\mathbf{n}$ transforms as the direct product representation $\Gamma_\mathbf{n} = \Gamma_N \otimes \Gamma_A$, where $\Gamma_N$ is a one-dimensional irrep describing the magnetic configuration. Consequently, $\mathbf{n}$ ($\Gamma_\mathbf{n}$) and $\boldsymbol{\sigma}_H$ ($\Gamma_A$) always belong to different representations, which naturally allows for their orthogonality.}

{This symmetry analysis bears a close resemblance to our discussion of magnetism-induced Raman scattering. In our framework, the magneto-Raman tensor $\boldsymbol{\alpha}^A$ transforms as $\chi_{\rm mag} \otimes \Gamma^{(r)}$, whereas the magnetic order parameter itself transforms as $\chi_{\rm mag}$. As a result, the  magneto-Raman vector ($\mathbf{J}$) and the magnetic order parameter ($\mathbf{m}$ or $\mathbf{n}$) often exhibit a similar orthogonal relationship.}

\end{spacing}
%%%%%%%%%%%%%%%%%%%%%%%%%%%%%%%%%%%%%
%\clearpage
%\bibliographystyle{apsrev4-1}
%\bibliography{Reflatex}
%%%%%%%%%%%%%%%%%%%%%%%%%%%%%%%%%%%%
%%%%%%%%%%%%%%%%%%%%%%%%%%%%%%%%%%%%%%%%%%%%%%%%%%%%%%%%%
%%%%%%%%%%%%%%%%%%%%%%%%%%%%%%%%%%%%%%%%%%%%%%%%%%%%%%%%%
\end{widetext}
\end{document}